\documentclass[11pt,a4paper]{article}
\usepackage{jcappub}
\usepackage[utf8]{inputenc}
\usepackage[table]{xcolor}
\usepackage{verbatim}
\usepackage{amsmath,amsfonts,amssymb}
\usepackage{subfloat}
\usepackage{float}
\usepackage{xcolor}
\usepackage{array, booktabs, tabularx, makecell, multirow}
\usepackage{multicol}
\usepackage{multirow}
\usepackage{tikz}
\usetikzlibrary{shapes.geometric, arrows}
\usepackage{tabularx}
\newcolumntype{s}{>{\hsize=.5\hsize}X}

\usepackage[misc]{ifsym} 

\tikzstyle{startstop} = [rectangle, rounded corners, minimum width=3cm, minimum height=1cm,text centered, draw=black, fill=red!30]
\tikzstyle{io} = [trapezium, trapezium left angle=70, trapezium right angle=110, minimum width=3cm, minimum height=1cm, text centered, draw=black, fill=blue!30]
\tikzstyle{process} = [rectangle, minimum width=3cm, minimum height=1cm, text centered, text width=3cm, draw=black, fill=orange!30]
\tikzstyle{decision} = [diamond, minimum width=3cm, minimum height=1cm, text centered, draw=black, fill=green!30]
\tikzstyle{arrow} = [thick,->,>=stealth]

\tikzstyle{lcdmparam} = [rectangle, rounded corners, minimum width=3cm, minimum height=1cm,text centered, draw=black, fill=yellow!30]
\tikzstyle{stockcode} = [rectangle, rounded corners, minimum width=3cm, minimum height=1cm,text centered, draw=black, fill=red!30]
\tikzstyle{modcode} = [rectangle, rounded corners, minimum width=3cm, minimum height=1cm,text centered, draw=black, fill=blue!30]
\tikzstyle{modparam} = [rectangle, rounded corners, minimum width=3cm, minimum height=1cm,text centered, draw=black, fill=orange!30]
\tikzstyle{goal} = [rectangle, rounded corners, minimum width=3cm, minimum height=1cm,text centered, draw=black, fill=green!50]

\usetikzlibrary{mindmap,backgrounds}
\usepackage{adjustbox}
\usepackage{tabularx}
\usepackage[normalem]{ulem}
\usepackage{soul}
\hypersetup{%
,urlcolor=red
,citecolor=red
,linkcolor=blue
}

\author[a]{Sankarshana Srinivasan,}
\author[b]{Shreya Prabhu,}
\author[a, c, d]{Kai Lehman,}
\author[e]{Ajiv Krishnan .V}
\author[a,d,f]{and Jochen Weller}

\affiliation[a]{Universit\"{a}ts-Sternwarte, Fakult\"{a}t f\"{u}r Physik, Ludwig-Maximilians Universit\"{a}t, Scheinerstraße 1, 81679 M\"{u}nchen, Germany}
\affiliation[b]{Indian Instititute for Science Education and Research (IISER), Thiruvananthapuram, Maruthamala P. O, Vithura, Kerala 695551, India}
\affiliation[c]{Center for Computational Astrophysics, Flatiron Institute, 162 5th Avenue, New York, NY, 10010, USA}
\affiliation[d]{Excellence Cluster ORIGINS, Boltzmannstr.~2, 85748 Garching, Germany}
\affiliation[e]{University of Sussex, Falmer, Brighton BN1 9RH, United Kingdom}
\affiliation[f]{Max-Planck-Institut f\"ur extraterrestrische Physik, Giessenbachstr.~1, 85748 Garching, Germany}

\emailAdd{ssriniva@usm.lmu.de}
\emailAdd{shreyaumeshprabhu@gmail.com}
\emailAdd{kai.lehman@physik.lmu.de}
\emailAdd{ajivkrishnan@gmail.com}
\emailAdd{jochen.weller@lmu.de}

\date{February, 2026}

\abstract{
As cosmology rapidly approaches the data-dominated phase of stage IV large scale structure surveys, the modelling of nonlinear scales has become a serious challenge that faces the community, particularly when analysing models beyond $w$CDM. In this work, we emulate the matter power spectrum in a phenomenological parameterisation of modified gravity in which a time-varying effective gravitational constant $\mu$ and a gravitational slip $\eta$ are binned in redshift. We are able to achieve accuracy $<1\%$ in the modified gravity boost relative to COLA (COmoving Lagrangian Acceleration) simulations, up to $k=1 h\,{\rm Mpc}^{-1}$. The emulator is publicly available in \href{https://github.com/sankarshana16/mg_binned_boost_emulator}{this git repo}. We forecast the constraining power for each bin using a simulated $3\times 2$pt LSST Y10-like data vector and a $6\times 2$pt LSST Y10 x Simons Observatory cosmic microwave background (CMB) lensing data vector. We recover the characteristic degeneracy between $\mu$ and $\eta$ previously identified in Fisher forecasts and demonstrate that the best-constrained direction corresponds to the combination $\Sigma=\mu(1+\eta)/2$ which governs the lensing potential. We show that while large scale structure is sensitive to growth of structure at low redshift, CMB lensing extends the sensitivity to a higher redshift range. These results demonstrate that fast emulation of nonlinear modified-gravity effects enables full Bayesian analyses of model-agnostic gravity parameterisations with realistic survey data vectors and astrophysical systematics. 
 }


\title{Cosmological gravity on all scales V: MCMC forecasts combining large scale structure and CMB lensing for binned phenomenological modified gravity}

\begin{document}

\maketitle

\section{Introduction}
One of the central challenges in the era of Stage~IV large scale structure surveys is the accurate modelling of cosmological observables on nonlinear scales. In the context of modified gravity, predicting nonlinear observables on a timescale that is viable in a cosmological data analysis pipeline is only possible for a select few models such as $f(R)$ gravity \cite{Bai_2024, Casares2023, Arnold2022, Ramachandra_2021}, the $n$DGP braneworld model of gravity \cite{Fiorini_2023, Ruan_2024}. As in $\Lambda$CDM, there is also a concerted effort to develop modelling frameworks that go beyond standard two point functions, with emulators for the higher order statistics \cite{Fiorini_2021, Hoyland_2025, orjuelaquintana2025}, or even field level emulators \cite{Saadeh_2024} being developed recently for $f(R)$ gravity. However, there is a severe lack of tools that can compute the nonlinear matter power spectrum that are validated by simulations for the vast majority of the model-space. This has led to severe scale cuts being employed in order ensure reliability \cite{DES_2019_extensions}.

It is this gap in the modelling that this body of work \cite{ref:DanPF, Srinivasan2021, Srinivasan_2024, Srinivasan_2025} has been developed to address. In the past iterations of this series, we have developed the formalism to model the nonlinear observables in modified gravity in a model-agnostic, data-driven context. Effectively, we parameterise the strength of clustering by a dimensionless re-scaling of the Poisson equation (the so-called $\mu$ parameter) and the gravitational slip (the $\eta$ parameter) as they appear in general relativity (GR). In particular, we  developed $N$-body simulations from which we computed the nonlinear matter power spectrum for different step-function like deviations from GR. The non-linear response to deviations in $\mu$ has been studied in the past through dedicated $N$-body simulations, \cite{ref:Cui1, ref:Cui2}, but in the limiting case where $\mu$ is constant over the entire simulation time-period. On the other hand, studies that forecast \cite{ref:Casas2017} (with simulated data) or infer (with real data) constraints on binned $(\mu, \eta)$ have been limited to exclusively linear modelling, with the Dark Energu Spectroscopic Instrument (DESI) collaboration recently reporting $\mathcal{O}(10)\%$ limits on purely time-binned $\mu$ \cite{Ishak_2025}. These constraints degrade when $\mu$ is also binned in scale, which shows that neglecting non-linear modes can severely limit the precision of future surveys.

In the past iterations of this work, simulations were used to validate an end-to-end pipeline that computes the full combination of 2-point functions of cosmic shear, galaxy clustering and their cross correlation, the so-called $3\times 2$pt data vector. We performed Fisher forecasts with this pipeline, that allowed one to forecast constraining power on the joint set of $\Lambda$CDM and modified gravity parameters. Similar pipelines have been adopted by the \textit{Euclid} collaboration to forecast general parameterisations of modified gravity \cite{Frusciante_2024, euclid_param_MG_forecast}.

However, this pipeline involved the running of the halo-model based \texttt{ReACT} code \cite{Cataneo_2019, BoseReACT, Bose_2021, Bose_2023}, which lacks the computational speed required to be directly implemented in a typical Bayesian Markov Chain Monte Carlo (MCMC) analysis pipeline (see appendix \ref{app:react} for more details on emulating \texttt{ReACT}). In this work, we develop an emulator of the nonlinear matter power spectrum that enables fast predictions of nonlinear clustering in the presence of phenomenological deviations from GR parameterised by the $\mu$ and $\eta$ functions. The emulator is trained on a suite of COmoving Lagrangian Acceleration (COLA) simulations and provide predictions for the nonlinear power spectrum as a function of cosmological parameters and the modified gravity parameters describing deviations from GR in discrete redshift bins. 

This emulator is implemented within a full cosmological inference pipeline based on the \texttt{CosmoSIS} framework. The pipeline computes the complete set of two-point correlation functions of large scale structure observables, including cosmic shear, galaxy clustering and galaxy-galaxy lensing, and incorporates modelling of key astrophysical systematics such as baryonic feedback, intrinsic alignments, galaxy bias and photometric redshift uncertainties. In addition, we extend the analysis to include CMB lensing, thereby constructing the full $6\times2$pt data vector that combines large scale structure probes with CMB lensing auto- and cross-correlations.

Using this framework, we perform forecasts for a Stage~IV survey configuration representative of the LSST Year~10 dataset, combined with CMB lensing measurements expected from future high-resolution CMB experiments such as the Simons Observatory. The analysis explores the constraining power of the data vector on deviations from GR in multiple redshift bins and investigates how the inclusion of CMB lensing improves constraints relative to the standard $3\times2$pt combination.

The structure of this paper is as follows. We present the modelling and forecasting methodology used in this work in sec.~\ref{sec:methods}, starting with a detailed discussion of our modified gravity parameterisation, its philosophical implications and the binning implementation that we follow in sec.~\ref{sec:MG_methods}. We then move on to the construction and validation of the Gaussian Process emulator for the nonlinear power spectrum within this framework. In sec.~\ref{sec:Pipeline} we present the full cosmological inference pipeline and modelling of astrophysical and observational systematics, including baryonic feedback in sec.~\ref{sec:Baryons}, cosmic shear nulling and analytic covariance modelling in sec.~ \ref{sec:Covariance_modelling}. Our results are presented in sec.~\ref{sec:results}, where we examine the constraints obtained from $3\times2$pt and $6\times2$pt data vectors and analyse the degeneracy structure of the modified gravity parameters. Finally, we summarise our conclusions and discuss future extensions of this framework in sec.~\ref{sec:conclusion}.

\section{Methods}\label{sec:methods}
Our forecast methodology involves the creation of synthetic $3\times 2$pt and $6\times 2$pt data vectors at a fiducial $\Lambda$CDM cosmology. We will then analyse this data vector with different input models, each corresponding to a departure from GR in a different, discrete redshift interval. Similar analyses have been carried for various specific modified gravity models in the literature \cite{tsedrik2024}. For each analysis, we fix the redshift bin in which the modified gravity is `switched on'. This allows us to investigate the relative constraining power for different probe combinations for different bin choices. We perform our Bayesian parameter inference using the modular cosmological pipeline \texttt{CosmoSIS} \cite{ref:ZuntzCosmosis}. The likelihood evaluation combines projected two-point statistics with theoretical predictions for modified gravity and nonlinear structure formation.

Screening has been studied extensively in the context of modified gravity (see for example \cite{Brax_2021} and references therein). We note that the impact of screening in our analysis, given our choice of scale cut is unlikely to be very strong. However, in the interest of conducting preliminary tests, we adopt a similar approach to screening as in the recent Euclid forecast \cite{euclid_param_MG_forecast}, i.e., the `superscreened' approach where the one smoothly transitions from the MG linear $P(k)$ to the $\Lambda$CDM nonlinear $P(k)$. This transition takes place in the quasi-linear regime. We present the posterior distribution for the screened case and its impact in appendix \ref{app:screening}. 

Sampling of the posterior distribution is performed using the nested sampling algorithm \texttt{Nautilus}, enabling robust exploration of non-Gaussian posteriors and efficient evidence estimation. Convergence is assessed by monitoring how the range of likelihood values explored by the sampler evolves during the run and by verifying that the resulting posterior contains a sufficiently large number of effective samples. For all of our production runs, we use a live points set $n_{\rm live} = 2000$, $n_{\rm update} = 1500$ and $f_{\rm live} = 0.2$. We verify that our results are stable with respect to these choices by re-running the sampler with alternative settings, varying the number of live points ($n_{\rm live}={3000,4000}$), the update frequency ($n_{\rm update}={1000,500}$), and the target live-point fraction ($f_{\rm live}={0.1,0.15}$). In all cases the recovered posterior distributions remain statistically consistent.  

\subsection{Modified gravity modelling}\label{sec:MG_methods}
We make use of the post-Friedmann formalism that was introduced in \cite{ref:Milillo} and developed to include nonlinear modified gravity in \cite{ref:DanPF}, focusing on the ``Parameterised Simple 1st Post-Friedmann'' (PS1PF) equations developed in \cite{ref:DanPF}. We parameterize the response of structure formation to gravity via two free functions $\mu$ and $\eta$ (the tilde implies Fourier transformed quantities)
\begin{eqnarray}
\frac{1}{c^2}k^2 \tilde{\phi}_{\rm P} & = & -\frac{1}{c^2}4\pi a^2 \bar{\rho} G_{\rm N}\mu(a)\tilde{\Delta} \,, \label{eq:MGParam} \\
\tilde{\psi}_{\rm P} & = & \eta(a)\tilde{\phi}_{\rm P} \, ,
 \end{eqnarray}
where $\bar{\rho}$ is the background density, $\tilde{\Delta} = \tilde{\delta} - \frac{\dot{a}}{a}\frac{3}{c^2k^2}ik_i \tilde{v}_i$ is the gauge-invariant density contrast in Fourier space, $G_{\rm N}$ is Newton's constant, $\phi_P$ and $\psi_P$ are the standard Newtonian gravitational potentials\footnote{$g_{00}=-1-\frac{2\phi_P}{c^2}$; $g_{ij}=a^2\delta_{ij}\left(1-\frac{2\psi_P}{c^2} \right)$.} (normally found to be equal in GR) and $\mu(a)$ is a dimensionless function of time (redshift/scale factor) representing a change to the strength of gravity, $\eta(a)$ is the dimensionless gravitational slip parameter, that influences the geodesics of photons. These equations are derived by expanding the FLRW metric to one step beyond the leading order, at which order the equations describe structure formation on all scales (see \cite{ref:DanPF} for details). A key consequence of this parameterisation is that the dynamics of massive particles are purely governed by $\mu(a)$, which can be probed in dark matter only $N$-body simulations. Since $\eta$ purely affects photon geodesics, it can be modelled in post-processing. It was shown in \cite{ref:DanPF} that this parameterisation may be adapted to any modified gravity model (by suitable definition of the functional forms of $\mu(a, k)$ and $\eta(a, k)$, which for the most general case would be scale-dependent) with a well-defined Newtonian limit and a sufficiently small vector potential on cosmological scales. From a PS1PF perspective, the work in \cite{ref:Hassani2019, ref:HassaniNBodyMG} can be interpreted as showing that specific choices in terms of models and screening mechanisms may be mapped on to specific functional forms for $\mu$ in this approach. 

In contrast, our approach is designed to be maximally model-agnostic, in the sense that our analysis is designed to search for generic deviations from GR, irrespective of model-space predictions. Such null-tests are powerful, as they can be useful in ruling out large portions of the parameter space in the case of a null detection. The general advantages of such approaches has been noted in recent works that adopt philosophically similar (albeit with methodological differences) parameterizations \cite{Sakr_2025, Zanoletti_2025}. We note that the choice of restricting our analysis to time dependence is not a requirement of the underlying framework that we are using. Indeed, we have shown \cite{Srinivasan2021, Srinivasan_2024} that there is considerable complex phenomenology in the purely time-dependent binned case that warrants a dedicated analysis before one introduces additional complexity in the form of scale-dependence. 

So far, we have developed the infrastructure for the case where $(\mu, \eta)$ are switched on in a single redshift bin. We leave the more complicated case where multiple redshift bins can all have  $(\mu, \eta)$ values different from unity to future work, as this would require a significantly larger simulation suite and substantial computational resources. We remark that this is a first step towards a more general analysis that allows multiple deviations from GR across the entire redshift range of a given survey. Care would have to be taken in such an analysis to ensure that the number of parameters that need to be sampled doesn't degrade constraints, an advantage of studying the individual bin case first. Studying individual bins also provides a clean way to isolate the redshift sensitivity of different probes before introducing additional degeneracies associated with simultaneous deviations across multiple bins.

As we have discussed in the previous papers in this series \cite{Srinivasan2021, Srinivasan_2024, Srinivasan_2025}, we choose the bin edges in such a way that cover the entire redshift range between $0 \leq z \leq 3.$ in five discrete bins. These bins are chosen such that the incremental $\Lambda$CDM growth is identical in all of them. These are shown in Table \ref{tab:bins}. In our MCMC chains, we vary $(\mu, \eta)$ in one of these bins in a run (by fixing the bin index). Therefore, we have 5 unique analyses, with $(\mu, \eta)$ varied in a different, fixed redshift interval in each case. For all cases, we compute the linear matter power spectrum using the `traditional binning' implementation within \texttt{isitGR} \cite{ Dossett_2011, ref:IshakBin, Garcia_Quintero_2019}

\begin{table}
    \centering
 
    \begin{tabular}{|c|c|}
       \hline
       Bin 1 & $0 \leq z < 0.43$ \\
       Bin 2 & $0.43 \leq z < 0.91$\\
       Bin 3 & $0.91 \leq z < 1.47$\\
       Bin 4 & $1.47 \leq z < 2.15$\\
       Bin 5 & $2.15 \leq z < 3.0$\\
       \hline
    \end{tabular}

    \caption{The redshift bins for $\mu(z)$ and $\eta(z)$ in this work. }
    \label{tab:bins}
\end{table}

\subsubsection{Emulating the matter power spectrum}\label{sec:emulation}
In order to compute the nonlinear power spectrum, we build an emulator for the modified gravity boost factor, defined to be the ratio 
\begin{equation}
    B(z, k) = \frac{P_{\rm MG}(z, k)}{P_{\Lambda \rm CDM}(z, k)} \, ,
\end{equation}
where the modified power spectra and the $\Lambda$CDM power spectra at different redshifts are measured from snapshots of COmoving Lagrangian Acceleration (COLA) simulations run on a modified version of the FML \cite{Winther_2023} library. The initial conditions are identical for all these simulations, which ensures that the cosmic variance error at low $k$ is nullified. The simulation suite is constructed using 500 latin hypercube samples over the cosmological and modified gravity parameters, as indicated in Table \ref{tab:Priors}. These simulations have $512^3$ particles in a $500\, h^{-1}\,{\rm Mpc}$ box. We verify the well-known result in the literature that the boost factor measured from these simulations matches the full $N$-body boost factor to within 1\% up to $k=1\,h\,{\rm Mpc}^{-1}$.

Given the high dimensionality of the power spectra, we employed a Reduced-Order Modeling approach. First, the training ensemble of COLA simulations was compressed using Principal Component Analysis (PCA). We retained the first five principal components, which account for the dominant variance in the matter distribution across the ensemble. A Gaussian Process (GP) regressor was then trained to map the primary cosmological parameters to these five PCA coefficients. By utilizing a RBF kernel with optimized length scales, the emulator captures the smooth, nonlinear dependencies of the cosmic web on the underlying dark energy and matter density parameters. This framework enables the rapid generation of power spectra, bypassing the computational overhead of running new COLA simulations for every point in the parameter space. The input parameters are $\{\Omega_{\rm m}, \Omega_{\rm b}, h, n_{\rm s}, A_{\rm s}, i, \mu, z_{\rm arr}\}$, where $i$ is the bin index (see table \ref{tab:bins} \footnote{We have discussed in the previous papers of this series that $\eta$ doesn't affect the geodesics of the DM particles, and therefore has no effect on $P(k)$. It is modelled at the level of the convergence power spectrum and is therefore not an input parameter for our emulator.} for the bin edges that correspond to different values of $i$) and $z_{\rm arr}$ is the array of redshift values (with the largest redshift possible being $z=5$). 

To motivate the use of the emulator, we compare its computational performance to the \texttt{ReACT} framework. For a typical likelihood evaluation within our analysis pipeline, the emulator achieves a runtime of $\mathcal{O}(0.5\,{\rm s})$ per evaluation. This corresponds to an approximate speedup factor of $\sim 90$ relative to the equivalent \texttt{ReACT}-based calculation. Such an acceleration is essential for performing full Markov Chain Monte Carlo analyses in high-dimensional parameter spaces, particularly when including multiple nuisance parameters and tomographic modified gravity bins. To motivate the use of the emulator, we compare its computational performance to the halo-model based \texttt{ReACT} framework \cite{Cataneo_2019, BoseReACT}. For a typical likelihood evaluation within our analysis pipeline, the emulator combined with the \texttt{BCemu} baryonic feedback model achieves a runtime of $\mathcal{O}(0.5\,{\rm s})$ per evaluation. This corresponds to an approximate speedup factor of $\sim 90$ relative to the equivalent \texttt{ReACT}-based calculation. Such an acceleration is essential for performing full Markov Chain Monte Carlo analyses in high-dimensional parameter spaces, particularly when including multiple nuisance parameters and tomographic modified gravity bins. We note that the  computational cost of GP regression increases with the dimensionality of the input parameter space. While this remains sufficiently efficient for the analyses considered here, future extensions involving pipelines with multiple MG redshift/scale bins simultaneously switched on, or simulation-based inference pipelines may benefit from replacing the Gaussian Process with feed-forward neural network architectures which can provide significantly faster inference once trained. 

\begin{figure}
    \centering
    \includegraphics[width=\linewidth]{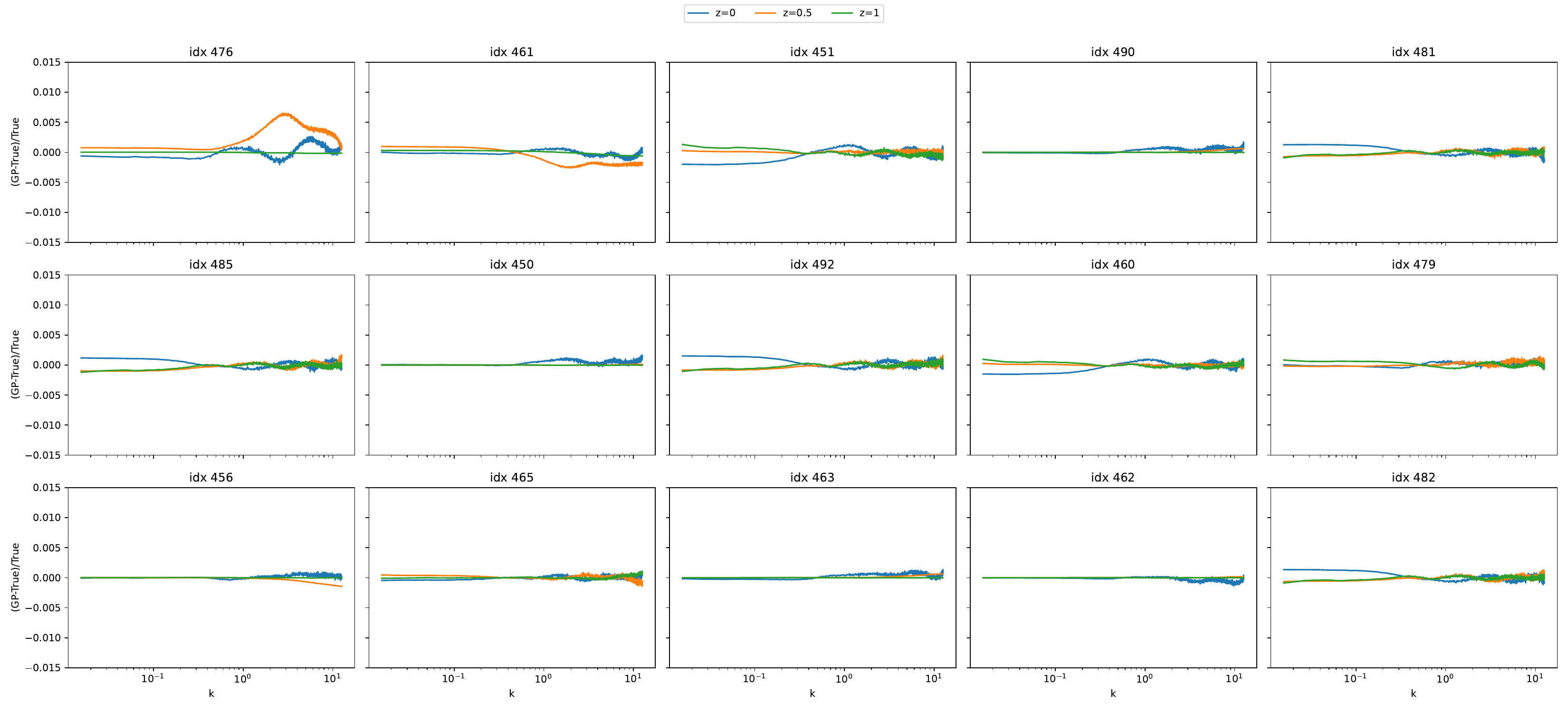}
    \caption{We show the fractional difference between the emulator prediction and the ground-truth simulation output for a subset of cosmologies from the validation set that were not used during training. The emulator accurately reproduces the boost across the full range of wavenumbers and redshifts considered, with deviations remaining small even in the nonlinear regime. We note that the accuracy is within 1\% across the entire validation set. This demonstrates that the emulator reliably interpolates the modified-gravity corrections across the parameter space relevant for the analysis.}
    \label{fig:pk_validation}
\end{figure}

\begin{figure}
    \centering
    \includegraphics[width=0.45\linewidth]{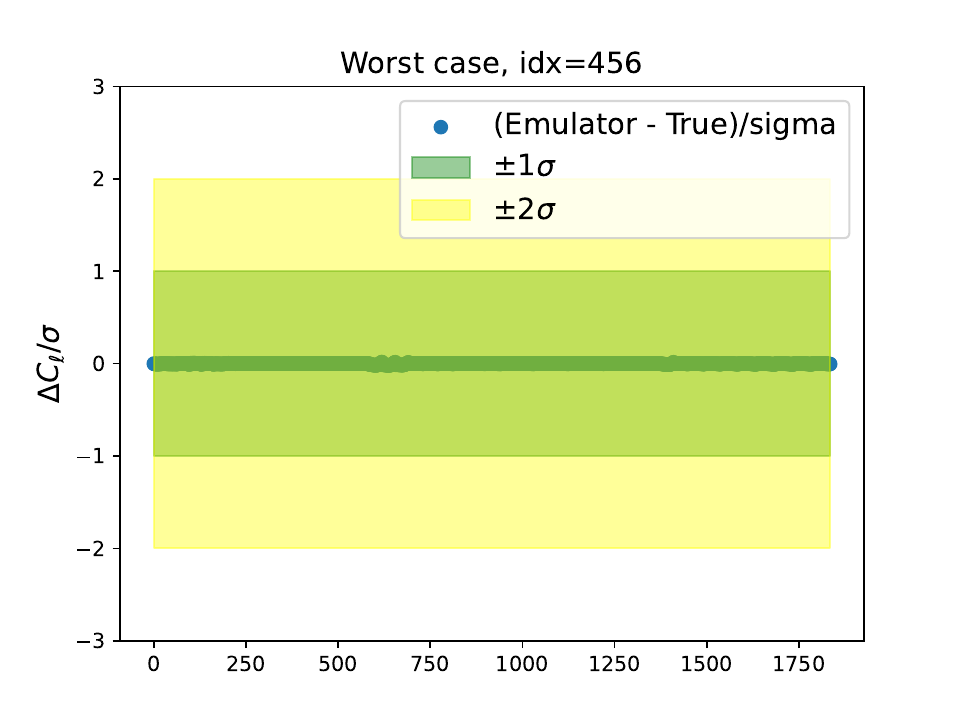}
    \includegraphics[width=0.45\linewidth]{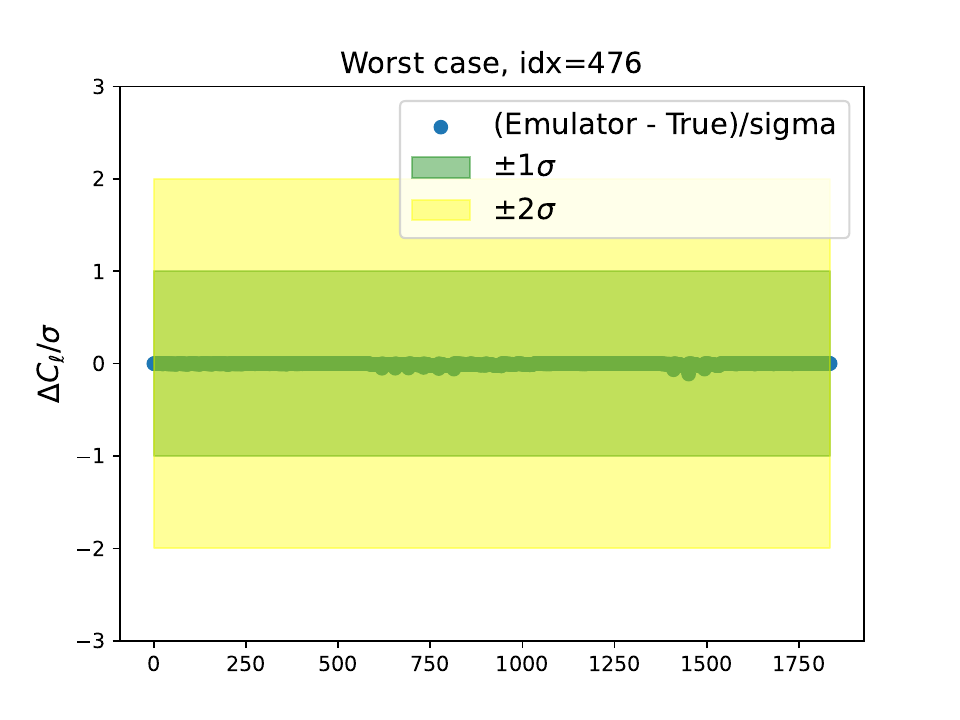}
    \caption{We show the difference between the data vector computed using the emulator and that obtained from the ground-truth simulations, normalized by the statistical uncertainty of the data vector. Results are shown for our two worst-case cosmologies from the validation set (i.e., the cases where the discrepancy between COLA and the emulator is maximum). The induced shifts in the observables remain well within the statistical uncertainty, demonstrating that emulator inaccuracies do not significantly affect the predicted data vector or the resulting parameter inference.}
    \label{fig:cl_validation}
\end{figure}

We explicitly validate the emulator performance at the level of the power spectrum, and the observable $C(\ell)$. We verify that the emulator is able to match the `ground truth' (which are the boosts measured from the COLA simulation suite) to within 1\% for the entire validation set. We show this in fig.~\ref{fig:pk_validation}. We further demonstrate in fig.~\ref{fig:cl_validation} that the error in the emulator prediction is well within the error budget as calculated from our covariance matrix for a representative set of edge cases. The emulator can be found in \href{https://github.com/sankarshana16/mg_binned_boost_emulator}{this git repo}. 

We note that our setup of fixed IC simulations has been used before to model 2-point statistics in various contexts before, but this will not suffice if one is interested in going beyond 2-point statistics \cite{Krause_2025, Gong_2024, lehman2026c3nn}. In particular, for field level approaches to this problem, one would sample different ICs for different $\Lambda$CDM-MG simulation pairs.   

\begin{table}[]
    \centering
    \begin{tabular}{c|c|c|c}
        Parameter & Fiducial & Lower bound & Upper bound  \\
        \hline
         $\Omega_{\rm m}$ & 0.3156 & 0.25 & 0.35\\
         $\Omega_{\rm b}$ & 0.0492 & 0.04 & 0.055 \\
         $h$ & 0.6727 & 0.65 & 0.73  \\
         $n_{\rm s}$ & 0.9645 & 0.95 & 1.0 \\
         $\ln 10^{10} A_{\rm s}$ & 3.0587 & 2.9960 & 3.091 \\
         $\mu$ & 1.0 & 0.9 & 1.1 \\
         $\eta$ & 1.0 & 0.9 & 1.1 \\ 
         \hline
    \end{tabular}
    \caption{The priors bounds for the parameters in the latin hypercube that was used to construct the COLA simulation suite. Priors are flat within these bounds.}
    \label{tab:Priors}
\end{table}

\subsection{Forward modelling the large scale structure and CMB lensing data vectors}\label{sec:Pipeline}

To construct the full $3\times 2$pt and $6\times 2$pt data vectors, one needs the redshift distribution of source and lens galaxies that contribute to the signal. We assume an LSST Y10-like survey, in which the source and lens redshift distributions are modelled using the well-known Smail formula \cite{ref:Smail1994} 
\begin{equation}\label{eq:smail}
 n(z) \propto \left(\frac{z}{z_0}\right)^{2}\exp\left[-\left(\frac{z}{z_0}\right)^{\beta}\right]\,.
\end{equation}
For LSST Y10, we use \cite{LSST_SRD} $\beta = \{0.68, 0.90\}$, $z_0 = \{0.11, 0.28\}$ and $n_{\rm gal} = \{27.0, 48\}\,{\rm arcmin}^{-2}$. We smooth the $n(z)$ kernels with a Gaussian that quantifies photo-$z$ uncertainty, which we set $\sigma_{\rm z} = 0.05 (1+z)$. We assume a sky fraction $f_{\rm sky} = 0.35$. We then use the Limber approximation \footnote{We use the \texttt{project\_2d} module in Cosmosis to compute all of our projected power spectra, including shear, clustering, CMB lensing and all cross-correlations. } to compute the projected signal for the $3\times 2$pt observables given by 
\begin{equation}\label{eq:Cl}
    C_{ij}^{XY}(\ell)=c\int_{z_{\rm min}}^{z_{\rm max}}\text{d}z\frac{W_i^X(z)W_j^Y(z)}{H(z)r^2(z)}P(k_{\ell},z)\, ,
\end{equation}
where $k_{\ell}=(\ell+1/2)/r(z)$, $r(z)$ represents the comoving distance as a function of the redshift, and $P(k_{\ell},z)$ stands for the nonlinear matter power spectrum evaluated at a wavenumber $k_{\ell}$ and redshift $z$. The kernels or window functions are given by
\begin{align}
    W_i^{\rm G}(k,z)=&b_i(k,z)\frac{n_i(z)}{\bar{n}}\frac{H(z)}{c}\, ,\\
     W_i^{\rm L}(k,z) =& \frac{3}{2}\Omega_{\rm m} \frac{H_0^2}{c^2}(1+z)\,r(z)\,\Sigma(z)
\int_z^{z_{\rm max}}{\text{d} z'\frac{n_i(z')}{\bar{n}_i}\frac{r(z'-z)}{r(z')}}\nonumber\\
  &+W^{\rm IA}_i(k,z)\, ,
\end{align}
where the `G' and `L' labels signify clustering and lensing, respectively and $\Sigma = \frac{1}{2}\mu \left[1+\eta \right]$. The ratio $n_i(z)/\bar{n}$ represents the normalised galaxy distribution as a function of redshift, while $b_i(k,z)$ is the galaxy bias in the $i$-th tomographic bin. Note that in our pipeline, we just vary the linear bias parameter in each bin assuming no additional scale/redshift dependence. The effects of modified gravity will be encapsulated in $\Sigma(z)$ for the lensing potential and the matter power spectrum $P(k_{\ell},z)$. The intrinsic alignment contribution to the weak lensing kernel enters through the $W_i^{\rm IA}(k,z)$ term. We consider the nonlinear alignment model with the single free parameter being the amplitude of intrinsic alignments, $A_{\rm IA}$ \cite{Kirk_2012_IA, Bridle_2007}. We remark that while this assumption may be overly simplistic, it is important to note that no detailed study of intrinsic alignment including higher order tidal terms in the context of modified gravity. We show in table \ref{tab:Systematic_Priors} the fiducial values and prior ranges for the systematic parameters that we marginalize over in our analysis. 

\subsubsection{CMB Lensing}
The CMB lensing power spectrum may be modelled using the same intuition as in weak-lensing, but with the cosmic microwave background being the background source, rather than galaxy populations. We make use of the Limber approximation to write \cite{LEWIS_2006} 
\begin{equation}\label{eq:CMB_lensing_convergence}
    C_{\ell}^{\rm CMB} = c \int_{z=0}^{z = z_{*}} dz \frac{W_{\rm CMB}^2}{H(z)r^2(z)} P(k_{\ell}, z) \,  
\end{equation}
where the CMB lensing kernel is given by
\begin{equation}\label{eq:CMB_lensing_kernel}
    W_{\rm CMB} = \frac{3}{2}\Omega_{\rm m} \frac{H_0^2}{c^2}(1+z)\,r(z)\,\Sigma(z)  \frac{r(z_*) - r(z)}{r(z_*)} \, ,    
\end{equation}
where $z_*$ is the redshift corresponding to recombination. It is worth remarking here that this kernel peaks at $z\sim 2$, a point which will help explain much of the results to come. 

We assume Simons Observatory-like error bars, where we make use of the publicly available forecast pipeline that provides the Large Aperture Telescope (LAT) lensing reconstruction noise curves \cite{Ade_2019} \footnote{\url{https://github.com/simonsobs/so_noise_models}}. The CMB lensing reconstruction assumes instrumental characteristics representative of the Simons Observatory large-aperture telescope. We adopt a Gaussian beam with full width at half maximum $\theta_{\rm FWHM}=1.4'$ corresponding to the 145\,GHz channel that dominates lensing
reconstruction forecasts. We cut our CMB lensing data vector at $\ell = 2000$. We will now describe our covariance modelling.  
\begin{table}[]
    \centering
    \begin{tabular}{c|c|c|c}
         Parameter & Fiducial & Lower bound & Upper bound \\
         \hline 
         NLA $A_{\rm IA}$ & 1.0 &  0.5 & 4.5  \\
         linear bias $b_k$ & 2.0 & 1.0 & 4.0 \\
         photo-z source bias $\varphi_j$ &  0.  & -0.01 & 0.01 \\
         photo-z lens bias $\vartheta_k$ & 0. & -0.01 & 0.01 \\
 \hline
    \end{tabular}
    \caption{The systematic parameters in our analysis that we marginalize over. Note that the index $j$ corresponds to the source bins, of which there are 5, while the index $k$ corresponds to the 10 lens bins. }
    \label{tab:Systematic_Priors}
\end{table}

\subsubsection{Modelling baryonic feedback}\label{sec:Baryons}
The problem of modelling baryonic feedback has received considerable attention in the context of $\Lambda$CDM, with multiple groups constructing complex hydrodynamic simulation suites capable of modelling a very wide range of feedback implementations \cite{Le_Brun_2014, McCarthy_2016, Springel_2017, Dave_2019, Villaescusa-Navarro_2021, Bird_2022, Schaye_2023}. Effectively, this corresponds to a very wide theoretical prior on the feedback parameters, which translates to a large uncertainty on the clustering and lensing signal in the nonlinear regime. This presents a unique problem for large scale structure, i.e., the nonlinear scales represent the majority of the volume of data-sets, but baryonic feedback dominates (to a large extent) the nonlinear regime. There is now a concerted effort within cosmology to infer the `correct' feedback parameters, or at the very least, enforce a `theory' informed prior on these parameters so as to retain some sensitivity to the nonlinear scales \cite{Eifler_2015_PCA_Baryons, Schneider_2019, Schneider1_2020, Schneider2_2020, xu2026constrainingbaryonicfeedbackcosmology, Lehman_2025, Bigwood_2024, Bigwood_2025, bigwood2025kineticsunyaevzeldovicheffect, Theis:2024mnr, Reischke:2023tqh, Reischke:2025srr,  wayland2026probingbaryonicfeedbackfast}. We will return to how we mitigate for baryonic feedback in this work later in this section. 

In order to model the effect of feedback, we employ the publicly available \texttt{BCemu} (Baryonic Correction Emulator \footnote{\url{https://github.com/sambit-giri/BCemu}}) which implements the Baryonification Model \cite{giri2021emulation}. The emulator computes a baryonic boost $S(k)$, while taking in the baryon-fraction as input, along with a three-parameter baryonification model. These parameters are the cut-off mass $M_{\rm c}$, the gas profile slope $d_{\eta}$, and the ejection factor $\theta_{\rm ej}$. In our MCMC chains, we marginalize over these three parameters. We show in table \ref{tab:Baryon_Priors} the fiducial value of these parameters and the prior ranges that we adopt in our analysis. An important caveat to note here is that we assume no correlation between the baryonification parameters and the modified gravity parameters. This effectively implies that $S(k)$ is not affected by variation of $\mu$ or $\eta$. This is a simplifying assumption and has indeed been demonstrated in a halo model context \cite{Mead_2016_baryon_MG_neutrinos}, but requires running computationally expensive hydrodynamic simulations with the binned approach taken here implemented in order to rigorously validate, which is beyond the scope of this work. It is worth noting that joint simulations of modified gravity and baryonic feedback have noticed that the two are not completely degenerate \cite{Mitchell_2019}, since modified gravity affects growth on all scales (impacting the linear $P(k)$ and halo abundance, while feedback only affects small-scale clustering and internal halo structure. Since we do not extend to scales that probe this ($k>1\,h\,{\rm Mpc}^{-1}$), we do not expect this assumption to have a strong impact on our results.   

\begin{table}[]
    \centering
    \begin{tabular}{c|c|c|c}
        Parameter & Fiducial & Lower bound & Upper bound  \\ 
        \hline 
        $M_{\rm c}$ & 13.5 & 11.0 & 15.0 \\ 
         $d_{\rm eta}$ & 0.097 & 0.05 & 4.0 \\ 
         $\theta_{\rm ej}$ & 4.7 & 2.0 & 8.0 \\ 
         \hline
    \end{tabular}
    \caption{The baryonic feedback parameters and their priors in the \texttt{BCemu} module that we marginalize over in our analysis. }
    \label{tab:Baryon_Priors}
\end{table}

\subsubsection{Covariance modelling and the BNT transform}\label{sec:Covariance_modelling}
We assume a Gaussian $3\times 2$pt covariance, which we compute at the fiducial $\Lambda$CDM cosmology, at which we also compute our simulated data vector. For two observables $A,B$ and $C,D$, the covariance between bandpowers at multipoles $\ell_1$ and $\ell_2$ is given by \cite{Fang_2021}
\begin{equation}
\mathrm{Cov}\!\left[
C^{ij}_{AB}(\ell_1), C^{kl}_{CD}(\ell_2)
\right]
=
\frac{\delta_{\ell_1 \ell_2}}
{f_{\rm sky}(2\ell_1+1)\Delta\ell}
\left[
\tilde{C}^{ik}_{AC}(\ell_1)\tilde{C}^{jl}_{BD}(\ell_2)
+
\tilde{C}^{il}_{AD}(\ell_1)\tilde{C}^{jk}_{BC}(\ell_2)
\right],
\label{eq:gaussian_cov}
\end{equation}
where $\delta_{\ell_1\ell_2}$ denotes the Kronecker delta, $f_{\rm sky}=0.35$ is the survey sky fraction, and $\Delta\ell$ is the multipole bin width. The observed spectra $\tilde{C}_\ell$ include the relevant noise contributions,
\begin{equation}
\tilde{C}^{ij}_{AB}(\ell)
=
C^{ij}_{AB}(\ell)
+
N^{ij}_{AB}(\ell).
\end{equation}
The noise terms are given by
\begin{align}
N^{ij}_{\kappa\kappa}(\ell) &= N^{\kappa\kappa}_{\rm CMB}(\ell), \\
N^{ij}_{\gamma\gamma}(\ell) &= 
\delta^{ij}\frac{\sigma_e^2}{n^{\,i}_{\rm source}}, \\
N^{ij}_{gg}(\ell) &= 
\delta^{ij}\frac{1}{n^{\,i}_{\rm lens}},
\end{align}
where $\sigma_e$ is the shape noise associated with the source galaxies, $n^{\,i}_{\rm source}$ and $n^{\,i}_{\rm lens}$ are the number densities of source and lens galaxies in bin $i$, and $N^{\kappa\kappa}_{\rm CMB}(\ell)$ denotes the CMB lensing reconstruction noise. In our analysis we adopt $\sigma_e = 0.26$ per shear component, consistent with the LSST DESC Science Requirements Document \cite{LSST_SRD}.

As we previously did in \cite{Srinivasan_2025}, we make use of the $k_{\rm cut}$ cosmic shear method \cite{Taylor_2018, Taylor_2021, Vazsonyi_2021, euclid_kcut}, which employs the Bernardeau-Nishimichi-Taruya (BNT) transform \cite{Bernardeau_2014} to linearly re-weight the cosmic shear data such that our confidence in the modelling of the power spectrum informs our scale cuts (for a detailed discussion on how the BNT transform works, and on the $k_{\rm cut}$ method, see section 2.3 in \cite{Srinivasan_2025}. See \cite{Barthelemy_2022} for a technical discussion on other applications of the BNT transform. For a general discussion on different techniques used for nulling, see e.g. \cite{piccirilli2025robustcosmicshearsmallscale}). We fix $k_{\rm cut} = 0.5\,h \, {\rm Mpc}^{-1}$ for all of our production runs. We remark that this is a conservative choice, motivated by the Fisher forecast analysis in the previous iteration of this series. 

The BNT transformation re-weights the shear signal such that each angular mode predominantly probes a well-defined range of wavenumbers, allowing a consistent implementation of the $k_{\rm cut}$ prescription described above. For observables involving galaxy clustering, the situation is different. The clustering kernel is comparatively narrow in redshift and therefore probes a more localized range of physical scales. For this reason we have made the universal choice of $k_{\rm cut}=0.5,h,{\rm Mpc}^{-1}$ to the clustering power spectra and to all cross-correlations involving galaxy clustering. Concretely, this prescription is used for galaxy clustering, galaxy–galaxy lensing, and the cross-correlation between galaxies and CMB lensing ($G\times\kappa_{\rm CMB}$). The CMB lensing auto-spectrum ($\kappa_{\rm CMB}\times\kappa_{\rm CMB}$) is treated analogously to the clustering observables, and for this case we employ a conservative cut of $\ell_{\rm max}=1500$. This hybrid approach follows standard practice in forecasting analyses and ensures that all observables entering the $6\times2$pt data vector are restricted to physical scales below $k_{\rm cut}=0.5,h,{\rm Mpc}^{-1}$. In practice, the weak-lensing sector provides the dominant contribution to the constraining power in most of the configurations considered here (particularly the strongest constrained principal component $\Sigma = \mu(1+\eta)/2$, see sec.~\ref{sec:results} for more detail), so the BNT-based $k_{\rm cut}$ prescription primarily governs the overall scale selection of the analysis.

To remain consistent with the BNT transformed data-vector, we transform the covariance matrix in the following way. 
Let $\hat{C}^{ef,gh}_{s_1,s_2}(\ell,\ell')$ denote the covariance between the angular power spectra $C^{ef}_{s_1}(\ell)$ and $C^{gh}_{s_2}(\ell')$, where $s_i \in \{ \mathrm{L}, \mathrm{G}, \mathrm{GGL} \}$ labels cosmic shear (L), galaxy clustering (G),
and galaxy–galaxy lensing (GGL), respectively. After applying the BNT transformation, the covariance becomes \cite{Vazsonyi_2021}
\begin{equation}
\tilde{C}^{ab,cd}_{s_1,s_2,\mathrm{BNT}}(\ell,\ell')
=
X^{a e b f}_{s_1}\,
X^{c g d h}_{s_2}\,
\hat{C}^{ef,gh}_{s_1,s_2}(\ell,\ell') ,
\label{eq:BNT_covariance}
\end{equation}
where repeated indices are summed over.

The transformation matrix $X^{a e b f}_{s}$ depends on the observable type:
\begin{equation}
X^{a e b f}_{s}
=
\begin{cases}
M^{a e} M^{b f}, & \text{if } s = \mathrm{L}, \\[6pt]
\delta^{a e} \delta^{b f}, & \text{if } s = \mathrm{G}, \\[6pt]
\delta^{a e} M^{b f}, & \text{if } s = \mathrm{GGL}.
\end{cases}
\label{eq:X_definition}
\end{equation}

Here:

\begin{itemize}
\item $M^{ab}$ is the BNT transformation matrix,
\item $\delta^{ab}$ is the Kronecker delta,
\item repeated indices are implicitly summed.
\end{itemize}
For the $6\times 2$pt case, we assume an identical setup, with the CMB lensing part of the data vector treated identical to how we treat the clustering data, i.e., we do not BNT transform the CMB lensing data. As in previous analyses, we check that our analysis pipeline obtains identical results to the case where the BNT transform is not applied when there are no scale cuts applied to the data vector. 

While non-Gaussian contributions from nonlinear structure formation are known to increase the covariance on small scales, the goal of this analysis is not to provide precision forecasts for a specific survey configuration. Instead, our objective is to investigate the sensitivity of large scale structure probes to phenomenological deviations from general relativity and to demonstrate that nonlinear modified-gravity modelling can be consistently incorporated within a full Bayesian inference pipeline. Since the analysis is restricted to scales below $k_{\rm cut} = 0.5\,h\,{\rm Mpc}^{-1}$, the Gaussian covariance provides a reasonable approximation for this purpose. We note that the Gaussianity assumption, coupled with the covariance only being evaluated at the fiducial cosmology, and computed with a fixed sky fraction $f_{\rm sky}=0.35$ for the full set of auto- and cross-spectra means that the improvement obtained when extending the analysis to the full $6\times2$pt data vector should be interpreted as somewhat optimistic. A more complete treatment including non-Gaussian covariance terms will be required for applications to real survey data and is left for future work.

\section{Results}\label{sec:results}

\begin{figure}
    \centering
    \includegraphics[width=0.9\linewidth]{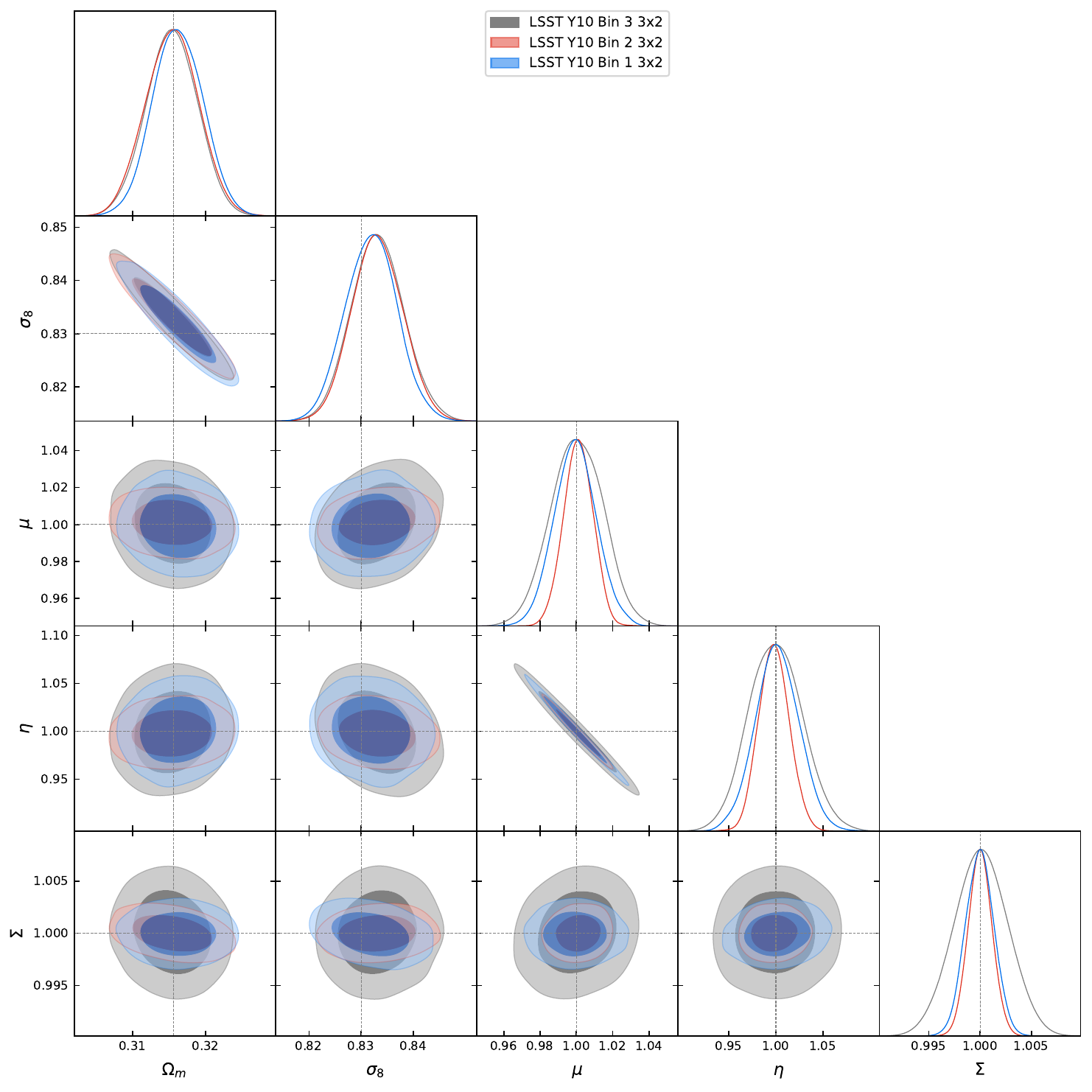}
    \caption{Posterior distributions for the cosmological and modified gravity parameters obtained from the $3\times2$pt analysis of the LSST Year~10 forecast for the first three redshift bins. The vertical and horizontal dashed lines indicate the fiducial $\Lambda$CDM values used to generate the mock data vector. In this analysis, we see that bin-2 is the redshift bin that yields the strongest constraints due to the higher signal-to-noise of the clustering and lensing observables and the larger contribution from nonlinear structure formation within the scales that we consider $k_{\rm cut}\leq 0.5 h\,{\rm Mpc}^{-1}$.}
    \label{fig:3x2_bin1_bin2_bin3}
\end{figure}

\begin{figure}
    \centering
    \includegraphics[width=0.9\linewidth]{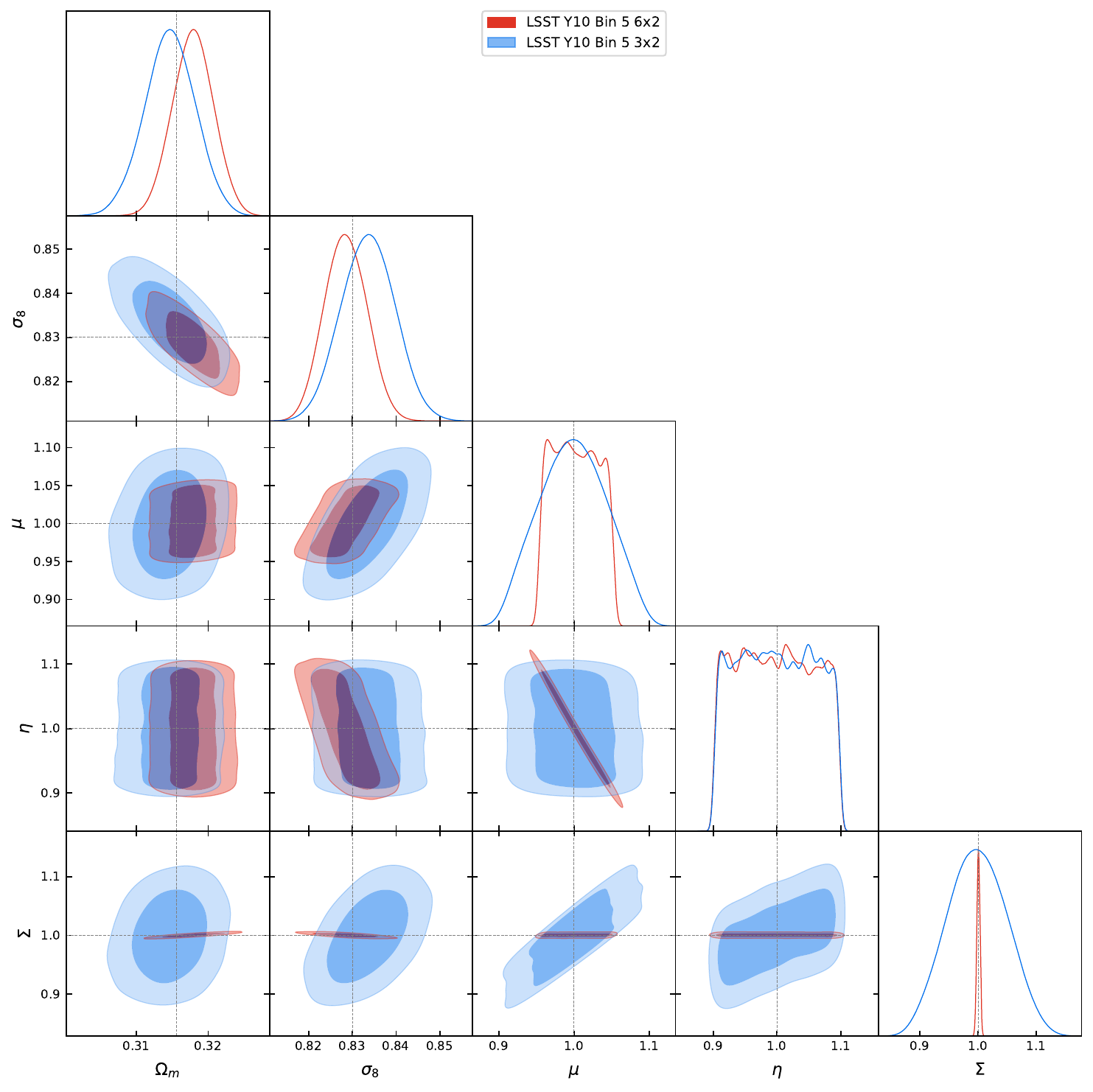}
    \caption{Comparison of the posterior constraints obtained from the $3\times2$pt (blue) and $6\times2$pt (red) data vectors for the highest redshift bin ($2.15<z<3$). The $6\times2$pt analysis additionally includes CMB lensing auto- and cross-correlations assuming Simons observatory-like survey specifications. The vertical and horizontal lines indicate the fiducial $\Lambda$CDM cosmology used to generate the mock data vector. The addition of CMB lensing tightens the constraint along the lensing-sensitive combination $\Sigma=\mu(1+\eta)/2$, reflecting the strong sensitivity of CMB lensing to the Weyl potential at high redshift.}
    \label{fig:6x2_bin5}
\end{figure}

\begin{figure}
    \centering
    \includegraphics[width=0.9\linewidth]{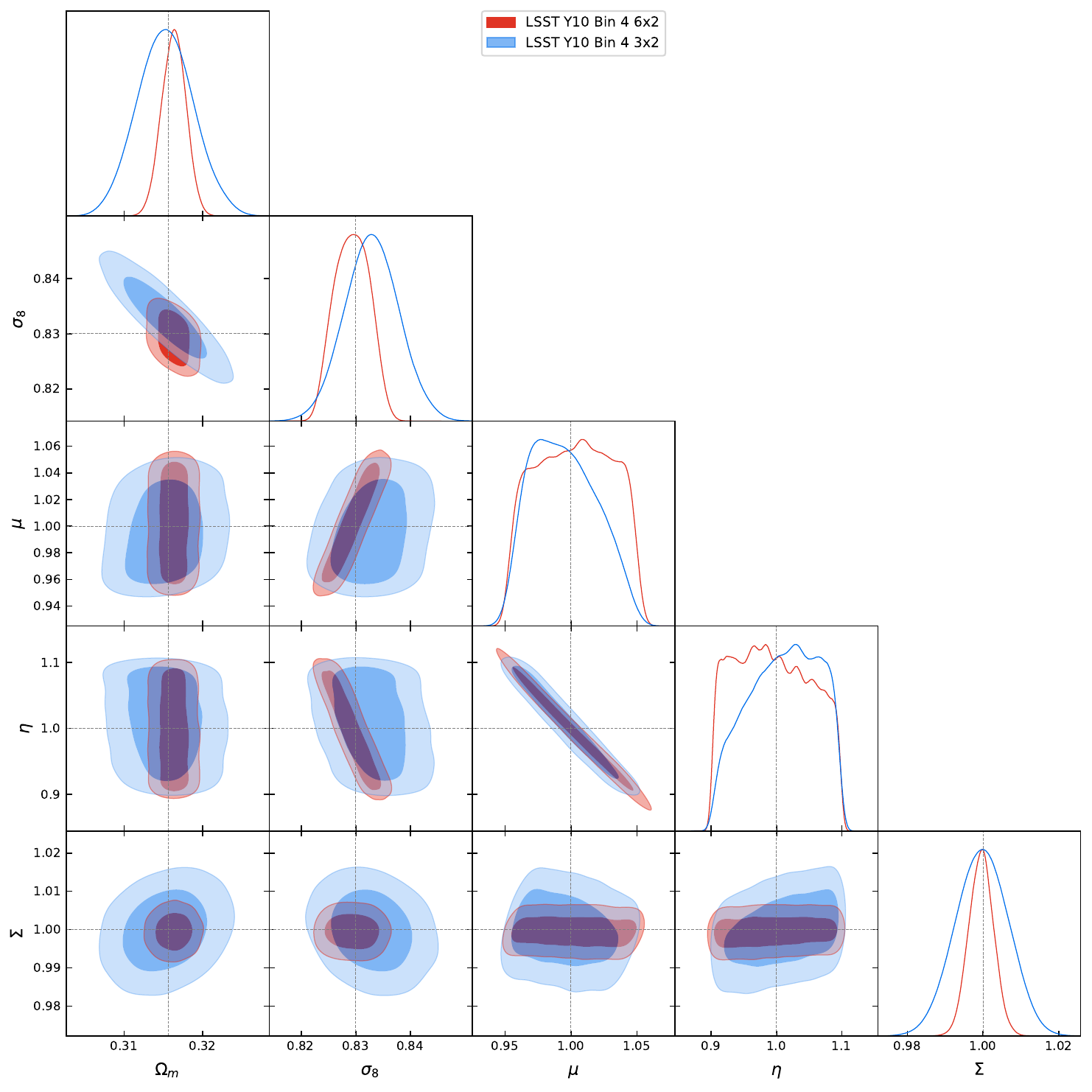}
    \caption{Identical as fig.~\ref{fig:6x2_bin5} but this time focusing on bin 4. We note that the gain is relatively smaller compared to the highest redshift bin (bin 5) due to the reduced sensitivity of CMB lensing to this bin.  }
    \label{fig:6x2_bin4}
\end{figure}

We present our results in stages. We begin by discussing the $3\times 2$pt forecast. We have already seen in previous Fisher forecasts \cite{Srinivasan_2025} that cosmic shear dominates the low redshift constraints. We have also seen a $\mu-\eta$ degeneracy that is difficult to disentangle even when one includes nonlinear information. We are able to reproduce these Fisher results in our MCMC chains, as can be seen in fig.~\ref{fig:3x2_bin1_bin2_bin3}. As previously noted, the lower redshift bins are constrained much more strongly by the combination of the $3\times 2$pt likelihood and synthetic data. In particular, we see that the constraining power on $\eta$ degrades substantially at $z>1$ (represented in our setup by bins 3, 4 and 5). Indeed, the sensitivity to the $\mu-\eta$ plane completely degrades in the case where our highest redshift bin (bin 5 in \ref{tab:bins}) is active. 

The problem of a lack of constraining power at higher redshift is somewhat alleviated by the addition of CMB lensing. This is shown explicitly in figs.\ref{fig:6x2_bin5} and \ref{fig:6x2_bin4}. This is an intuitive result, since the CMB lensing kernel is most sensitive to structure formation at $z\sim 2$, the CMB lensing kernel has significant overlap with the redshift bin(s) in which the modified gravity parameters are varied.  

One of our most striking and non-intuitive results is the constraining power associated to the $\Sigma = \mu(1 + \eta)/2$ parameter. In principle, this is the quantity that appears explicitly in the lensing likelihood (see eqs. \eqref{eq:Cl} and \eqref{eq:CMB_lensing_kernel}). However, it is important to reconcile the difference in uncertainty on $\mu$ and $\eta$, relative to $\Sigma$ and ensure that this gain is not due to any systematic or modelling inconsistency. In order to this, we carry out a principal component analysis on the $\mu-\eta$ posterior samples. The PCA is computed from the
weighted covariance matrix of $(\mu,\eta)$ obtained from the nested sampling chains.

\begin{figure}
    \centering
    \includegraphics[width=0.45\linewidth]{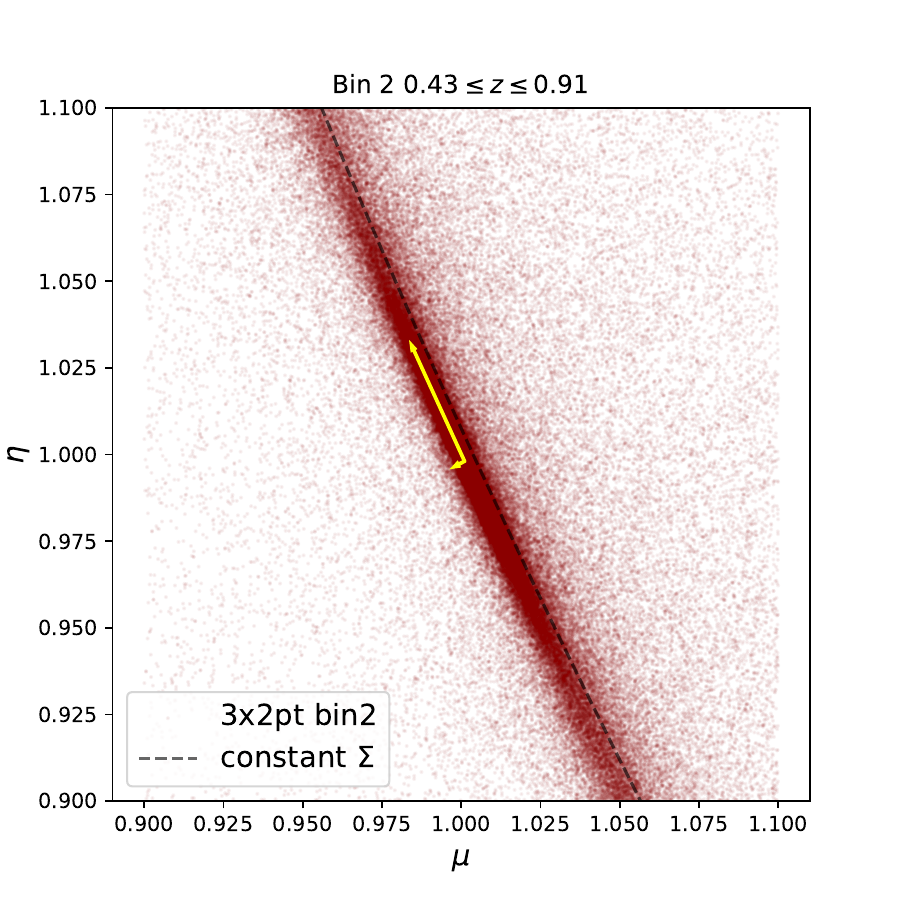}
    \includegraphics[width=0.45\linewidth]{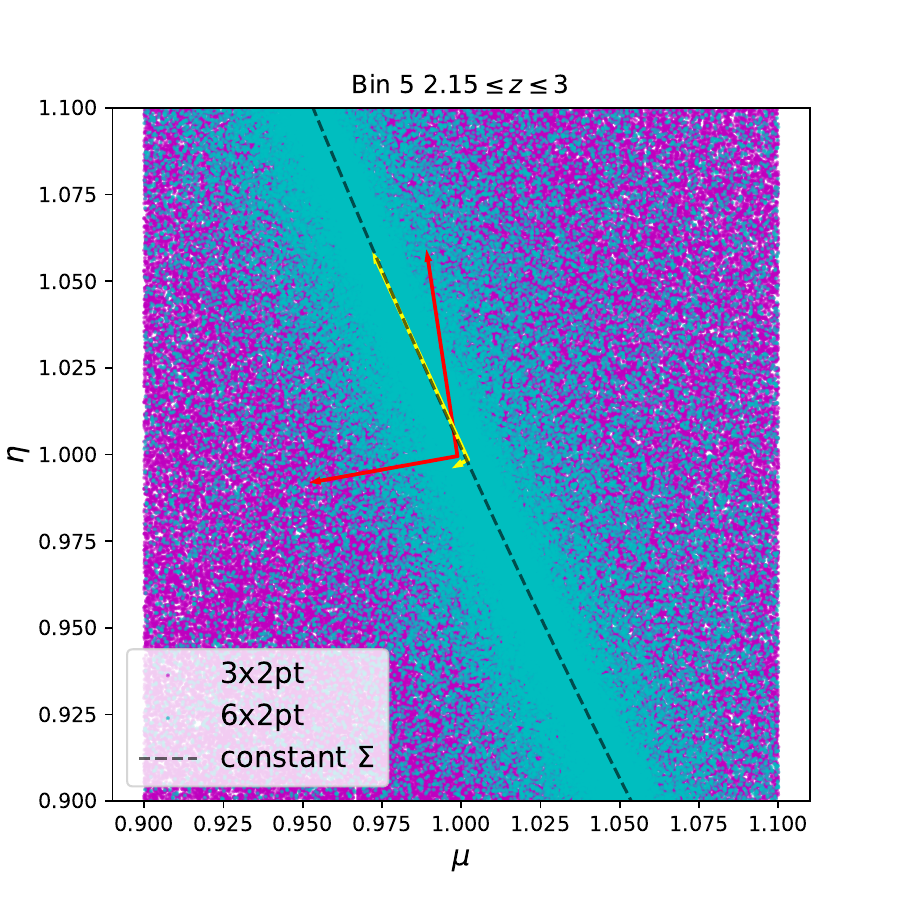}
    \caption{Principal component analysis of the $\mu - \eta$ posterior distributions. The left panel shows the posterior samples for the best constrained redshift bin obtained from the $3\times2$pt analysis (red points), with the corresponding PCA eigenmodes overplotted in yellow. The right panel shows the same analysis for the highest redshift bin, comparing the $3\times2$pt constraints (magenta points) with the $6\times2$pt constraints that include CMB lensing (cyan points). The principal component directions for the $3\times2$pt and $6\times2$pt analyses are indicated by the red and yellow lines respectively. The wide eigenmode corresponds to the well-known degeneracy between $\mu$ and $\eta$, while the tightly constrained direction closely aligns with the lensing-sensitive combination $\Sigma = \mu(1+\eta)/2$ which governs the amplitude of lensing observables. Note that the black dashed line indicates the line of constant $\Sigma$ in both panels. The inclusion of CMB lensing primarily tightens the constraint along this $\Sigma$ direction at high redshift.}
    \label{fig:PCA}
\end{figure}

For the best constrained redshift bin in the $3\times2$pt analysis, the parameter covariance matrix exhibits two orthogonal eigenmodes with different variances. The first mode corresponds to a broad degeneracy direction with standard deviation $\sigma_{\rm wide}\simeq0.005$, while the second mode is much more tightly constrained with $\sigma_{\rm tight}\simeq 0.001$. The factor of $\sim 5$ difference between these variances indicates that the likelihood effectively constrains combination of $\mu$ and $\eta$, while allowing for some freedom along the orthogonal direction. The tightly constrained eigenmode corresponds closely to $\Sigma$, which governs the amplitude of the Weyl potential responsible for gravitational lensing. This clearly indicates that the data constrain $\Sigma$ very tightly, while remaining relatively insensitive to orthogonal combinations of $(\mu,\eta)$. This behaviour is illustrated in the left panel of Fig.~\ref{fig:PCA}, where the posterior samples in the $(\mu,\eta)$ plane show a clear degeneracy direction along which $\Sigma$ remains approximately constant. The PCA eigenvectors align with this structure, with the tight eigenmode corresponding to variations perpendicular to the constant-$\Sigma$ direction and the wide eigenmode corresponding to motion along the degeneracy.

We perform the same PCA analysis for the highest redshift bin ($2.15<z<3$), where the addition of CMB lensing information is expected to have the largest impact. The $3\times2$pt analysis the covariance matrix again exhibits similar qualitative behaviour as in bin 1. However, with both posterior on $\mu$ and $\eta$ running into the prior boundary, the increased sensitivity to $\Sigma$ doesn't translate to a tight constraint on $\Sigma$. Including CMB lensing ($6\times2$pt) changes this significantly. While the wide degeneracy direction remains largely unchanged the orthogonal eigenmode becomes substantially more tightly constrained. This behaviour can be understood from the fact that both galaxy lensing and CMB lensing probe the Weyl potential and therefore directly constrain $\Sigma$. Consequently, the addition of CMB lensing data provides a direct probe of $\Sigma$ in this bin and significantly tightens the constraint on the eigenmode orthogonal to the $(\mu,\eta)$ degeneracy direction. This is consistent with the behaviour observed in the right panel of fig.~\ref{fig:PCA}, where the posterior samples show not only a clear improvement in the standard deviation of the posterior samples in the $\mu-\eta$ plane due to the addition of CMB lensing, but also highlights that the blue PCA eigenmodes (which correspond to the $6\times 2$pt analysis) are exactly along and orthogonal to the $\mu-\eta$ degeneracy direction along which $\Sigma$ remains approximately constant. We provide a summary of all of the constraints on our MG parameters in the various cases considered in this work in table \ref{tab:mg_constraints}.

\begin{figure}
    \centering
    \includegraphics[width=0.45\linewidth]{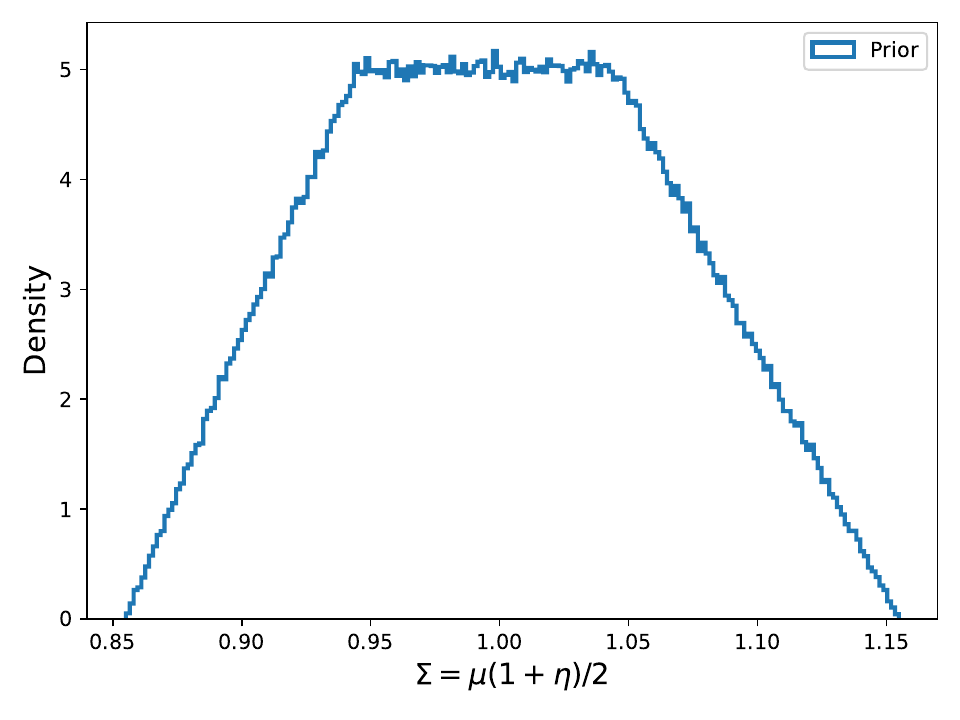}
    \includegraphics[width=0.45\linewidth]{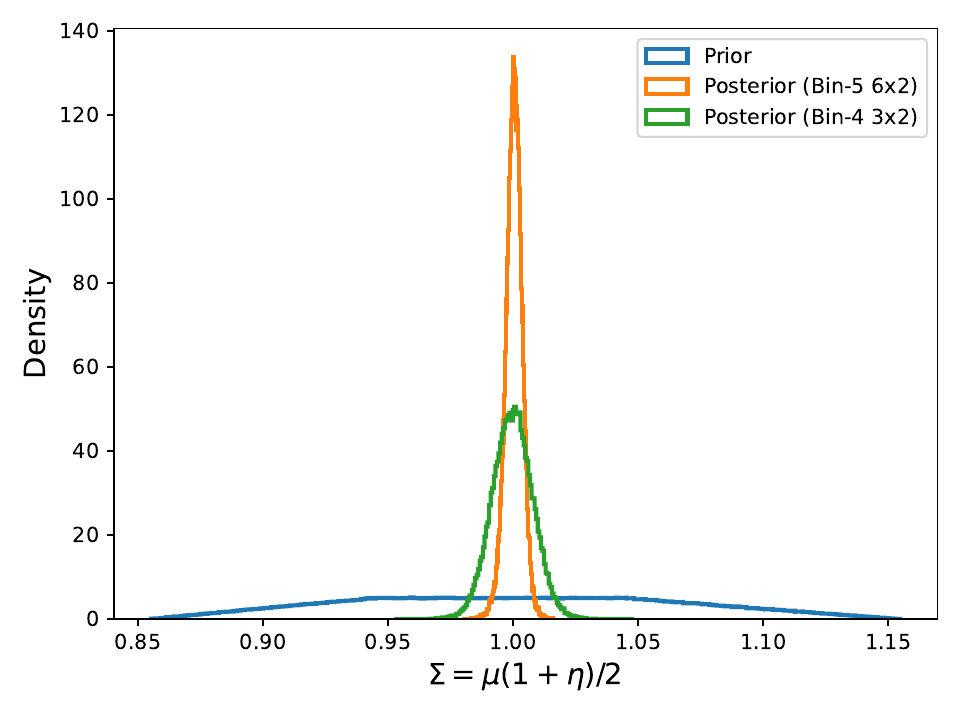}
    \caption{We show (in the left panel) the induced prior on $\Sigma$. The prior is obtained by sampling the uniform priors on $\mu$ and $\eta$ used in the analysis. Due to the nonlinear transformation between parameters, the resulting prior on $\Sigma$ is not uniform and peaks around the GR value of $\Sigma = 1$. In the right panel, we show the posterior from $6\times 2$pt analysis with Bin-5 in orange and the $3\times 2$pt analysis with Bin-4 in green. Both these posteriors are well within this prior range, demonstrating the robustness of the $\Sigma$ constraint. }
    \label{fig:Sigma_prior}
\end{figure}

\begin{table}[t]
\centering
\begin{tabular}{c c c c c}
\hline
Bin & Data vector & $\mu$ & $\eta$ & $\Sigma$ \\
\hline
1 & $3\times2$pt & $1.00 \pm 0.020$ & $1.00 \pm 0.045$ & $1.00 \pm 0.005$ \\
1 & $6\times2$pt & $1.00 \pm 0.020$ & $1.00 \pm 0.045$ & $1.00 \pm 0.004$ \\

2 & $3\times2$pt & $1.00 \pm 0.012$ & $1.00 \pm 0.025$ & $1.00 \pm 0.003$ \\
2 & $6\times2$pt & $1.00 \pm 0.015$ & $1.00 \pm 0.025$ & $1.00 \pm 0.003$ \\

3 & $3\times2$pt & $1.00 \pm 0.030$ & $1.00 \pm 0.048$ & $1.00 \pm 0.005$ \\
3 & $6\times2$pt & $1.00 \pm 0.035$ & $1.00 \pm 0.05$ & $1.00 \pm 0.005$ \\

4 & $3\times2$pt & $1.00 \pm 0.04$ & Prior-Dominated & $1.00 \pm 0.015$ \\
4 & $6\times2$pt & $1.00 \pm 0.06$ & Prior-Dominated & $1.00 \pm 0.009$ \\

5 & $3\times2$pt & Prior-Dominated & Prior-Dominated & Prior-Dominated \\
5 & $6\times2$pt & Prior-Dominated & Prior-Dominated & $1.00 \pm 0.009$ \\
\hline
\end{tabular}
\caption{
Marginalized $1\sigma$ constraints on the modified gravity parameters $\mu$, $\eta$, and $\Sigma=\mu(1+\eta)/2$ in each redshift bin for the $3\times2$pt and $6\times2$pt analyses. The addition of CMB lensing information in the $6\times2$pt case improves the higher redshift constraints on $\Sigma$, as discussed in the  text. 
}
\label{tab:mg_constraints}
\end{table}

Finally, we note that since we sample in $\mu-\eta$ space, the prior on $\Sigma$ is not uniform. We show in fig.~\ref{fig:Sigma_prior} the shape of this prior relative to the posterior (in bins 4 and 5). In particular, we emphasize that at $z>1.5$, $\Sigma$ is the quantity that the data vector and likelihood are most sensitive to, as compared to $\mu$ and $\eta$. This is an important point, and it might be more worthwile for future analyses to sample directly in $\mu-\Sigma$ space.

\section{Conclusion}\label{sec:conclusion}

In this work we have presented a forecasting framework for constraining phenomenological deviations from General Relativity using large scale structure observables in the context of upcoming Stage~IV surveys. Building on our previous work, which developed the formalism and validated the modelling of nonlinear modified gravity observables, we have constructed a full Bayesian inference pipeline capable of performing Markov Chain Monte Carlo forecasts for our model-agnostic parameterisation of gravity.

We introduce a fast emulator based on Gaussian Processes for the nonlinear matter power spectrum in the presence of step-like deviations from GR parameterised by the functions $\mu$ and $\eta$ in discrete redshift bins. The emulator is trained on pairs of matched COLA simulations in which modified gravity and $\Lambda$CDM runs share identical initial conditions, allowing us to emulate the nonlinear boost relative to the $\Lambda$CDM power spectrum. This approach enables rapid predictions of nonlinear clustering within a realistic cosmological inference pipeline.

We implement this emulator within a \texttt{CosmoSIS} pipeline that models the $3\times2$pt data vector consisting of cosmic shear, galaxy clustering and galaxy-galaxy lensing in harmonic space, and the $6\time2$pt data vector that includes CMB lensing and its cross-correlations with the aforementioned large scale structure observables. The pipeline includes modelling of astrophysical and observational systematics including baryonic feedback (with the \texttt{BCemu} emulator), intrinsic alignments, galaxy bias and photometric redshift uncertainties, as well as the application of BNT nulling to mitigate sensitivity to poorly modelled small-scale modes. To our knowledge, this represents the first model-agnostic modified gravity forecast performed using a full MCMC analysis that simultaneously incorporates baryonic feedback modelling and BNT nulling.

We additionally investigated the use of the halo model reaction approach implemented in \texttt{ReACT} as a possible fast prediction scheme for nonlinear modified gravity observables. While we were able to construct an emulator for the \texttt{ReACT} predictions, we found that the reaction method exhibits unstable behaviour away from the central region of the parameter prior volume in this phenomenological parameterisation, which can artificially inflate the inferred constraining power (see appendix \ref{app:react}). We therefore conclude that further development of the reaction framework is required before it can be reliably applied to model-agnostic modified gravity analyses of this kind.

From a computational perspective, the emulator developed in this work provides an approximate factor of $\sim 90$ speedup relative to equivalent \texttt{ReACT}-based calculations, enabling efficient full Bayesian analyses in high-dimensional modified gravity parameter spaces. While the present implementation employs Gaussian Process regression, future extensions to larger tomographic parameter spaces and simulation-based inference pipelines may benefit from neural-network based emulators with improved scaling properties.

Our forecasts adopt conservative scale cuts of $k_{\rm max}=0.5\,h\,{\rm Mpc}^{-1}$ in order to avoid poorly modelled nonlinear scales. As a result, screening mechanisms that operate predominantly at smaller physical scales are not explicitly modelled in this analysis. While screening effects become important for $k\gtrsim1\,h\,{\rm Mpc}^{-1}$, these scales lie beyond the range used in the present work. Incorporating screening physics will become increasingly important in future analyses that aim to extract information from smaller scales with higher order summary statistics or from field-level inference techniques.

The forecast results show that the constraining power of the $3\times2$pt data vector is strongest in the lowest redshift bin and decreases steadily towards higher redshift. This behaviour is consistent with the Fisher forecasts presented in our previous work and reflects the reduced nonlinear sensitivity and therefore higher susceptibility to degeneracies at higher redshift. A principal component analysis of the posterior distributions reveals that the data are primarily sensitive to the parameter combination $\Sigma = \frac{\mu(1+\eta)}{2}$, which governs the amplitude of lensing observables. The orthogonal combination of $\mu$ and $\eta$ remains weakly constrained, giving rise to the characteristic degeneracy observed in the posterior distributions.

Broadly, our results are consistent with previous phenomenological modified gravity studies that find lensing observables to be primarily sensitive to the Weyl-potential combination $\Sigma = \mu(1+\eta)/2$ (see \cite{Ishak_2025}, where the constraint on $\Sigma$ is $\sim 3$ stronger than that on $\mu$ due to combining with weak-lensing data from the Dark Energy Survey). The present work extends these analyses into the nonlinear regime within a full Markov Chain Monte Carlo framework, demonstrating that nonlinear information and multi-probe analyses will play an important role in future cosmological tests of gravity.

At higher redshifts, the $3\times2$pt observables alone provide relatively weak constraints on deviations from GR. The addition of CMB lensing significantly improves constraints in the highest redshift bins, particularly for $z>2$, by providing sensitivity to the integrated gravitational potential along the line of sight. This behaviour is clearly reflected in the principal component analysis, which shows that the inclusion of CMB lensing primarily tightens constraints along the $\Sigma$ direction. Additional validation tests of the  data vector are presented in appendix \ref{app:data_vector}.

Finally, we also study the impact of screening on our parameter constraints in \ref{app:screening}. We find that key impact of screening the matter power spectrum is in the $\mu-\eta$ degeneracy breaking. However, we still retain considerable constraining power on $\mu$ ($<5\%$) and $\Sigma$ ($<1\%$) across bins 1-3, even with screening implemented.  

The ultimate goal of this research is to build a comprehensive simulation suite, capable of testing deviations from GR in arbitrary combinations of redshift bins with a fully nonlinear cosmological inference framework. The emulator developed in this work represents a first step towards that objective, providing a fast and flexible tool for exploring phenomenological modified gravity models in the era of precision large scale structure surveys.

\section{Acknowledgements}
Sankarshana Srinivasan is grateful to Steffen Hagstotz, Daniel Thomas, Peter Taylor for useful advice and comments. Sankarshana Srinivasan was supported by an Alexander von Humboldt fellowship grant. KL acknowledges support via the KISS consortium (05D23WM1) funded by the German Federal Ministry of Education and Research BMBF in the ErUM-Data action plan. KL acknowledges support from Simons Foundation Pre-Doctoral program. 

\bibliography{ref.bib}
\bibliographystyle{apsrev4-1}

\appendix 

\section{\texttt{ReACT} emulator}\label{app:react}
The halo model reaction formalism \cite{Cataneo_2019, BoseReACT, Bose_2021} is based on a modified version of the halo model, and has been successfully used to accurately predict the matter power spectrum in a variety of $\Lambda$CDM extensions \cite{Bose_2023, euclid_param_MG_forecast}. Indeed, the \texttt{ReACT} code was validated against $N$-body simulations for the binned parameterisation considered here \cite{Srinivasan_2024}. However, this validation was done as a function of $\mu$ for fixed cosmological parameters. In the case where one is interested in computing $P(k)$ for arbitrary cosmologies within the prior range considered in this work (see table \ref{tab:Priors}), the code systematically fails for certain input cosmologies. This is due to the fact that spherical collapse recipe that the formalism follows is not viable for $\sigma_8$ values that are too far away from the fiducial. Essentially, the spherical collapse solver fails to produce physically meaningful halo collapse solutions. 

\begin{figure}
    \centering
    \includegraphics[width=0.5\linewidth]{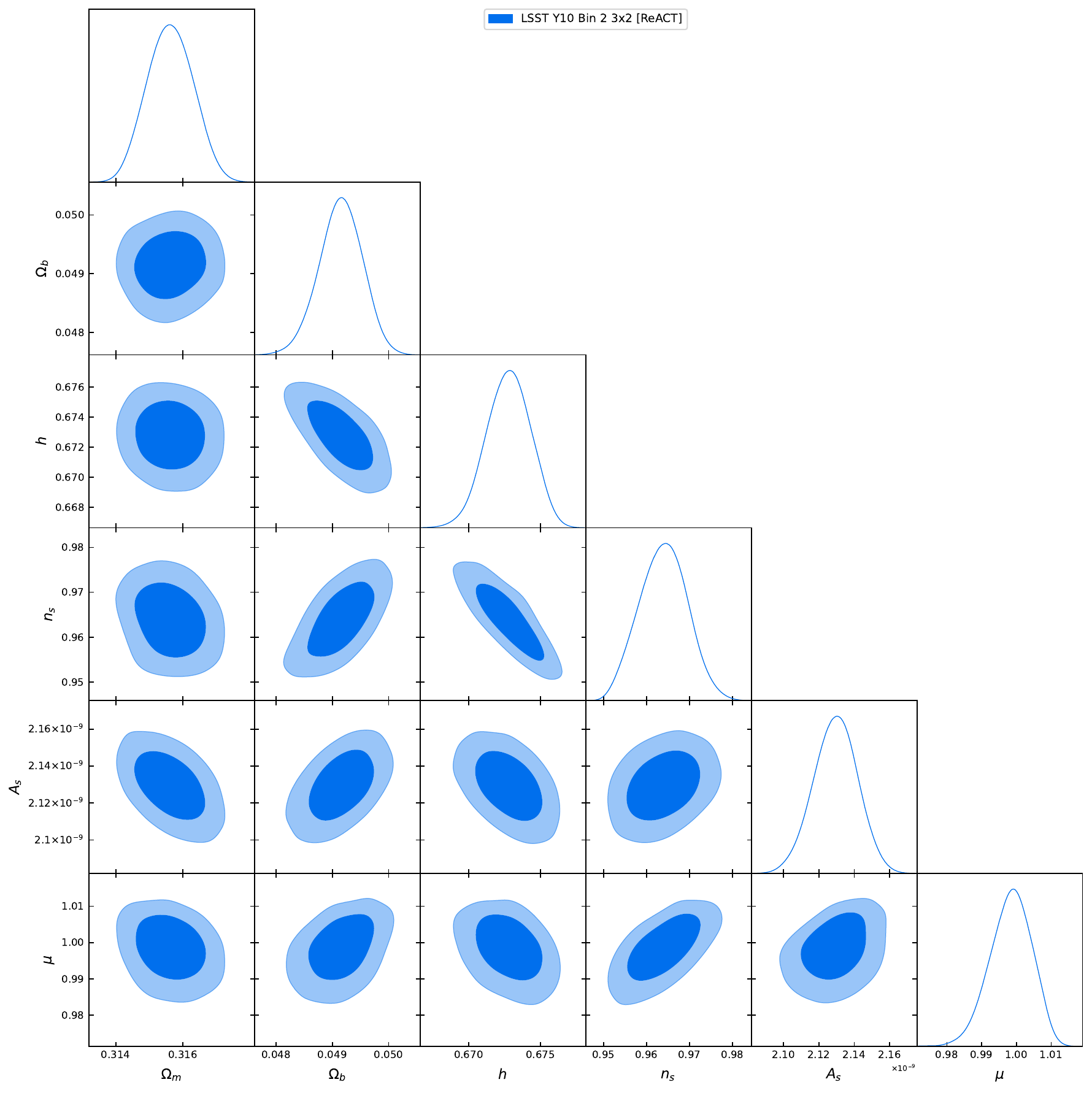}
 \caption{ Forecast constraints obtained using an emulator trained on the halo model reaction formalism implemented in \texttt{ReACT}. The triangle plot shows posterior constraints on a subset of cosmological and modified gravity parameters for an LSST Y10 $3\times2$pt analysis. While the constraint on the modified gravity parameter $\mu$ is broadly consistent with that obtained using the Gaussian Process emulator, several $\Lambda$CDM parameters appear artificially overconstrained. This behaviour arises because the \texttt{ReACT} predictions fail for a non-negligible fraction of the prior volume at extreme values of $\sigma_8$, leading to gaps in the training data and biasing the emulator response near the edges of  parameter space. As a result, the emulator spuriously suppresses parameter variations in regions where reliable training samples are unavailable, producing unrealistically tight posterior constraints.}
\label{fig:react_contours}
\end{figure}

We briefly describe our neural network architecture used for the training of the emulator and how the failure of the code affects the proceeding analysis.  The emulator is built upon a Feed-Forward Neural Network (FFNN) architecture, implemented using the PyTorch framework. To capture the non-linear boost (the so-called reaction $R(k)$), the model utilizes a deep configuration consisting of three hidden layers, with each layer containing 1,024 neurons. The network parameters were optimized using the Adam (Adaptive Moment Estimation) algorithm, chosen for its efficiency in handling sparse gradients and adaptive learning rate capabilities. The initial learning rate was set to $\alpha \approx 5.23\times10^{-4}$, determined via prior hyperparameter optimization. The training objective was to minimize the Mean Squared Error (MSE) loss function. 

We aimed to produce $10^6$ samples, but the code failed for about 15\% of the prior volume (see above discussion). Therefore, there is a lack of adequate training data at these extreme values of $\sigma_8$ (which corresponds to a subspace of the 8-dimensional space of input parameters). In this subspace, the emulator predictions are far away from the truth, resulting in artificially inflated constraining power (see fig.~\ref{fig:react_contours}). Despite the challenges, we find that the 1-dimensional posterior on $\mu$ for LSST Y10 (bin 2) is quite similar to that obtained by our Gaussian Processes emulator. However, the results are clearly overestimating constraining power on several $\Lambda$CDM parameters.

\section{Data vector validation}\label{app:data_vector}

\begin{figure}
    \centering
    \includegraphics[width=0.45\linewidth]{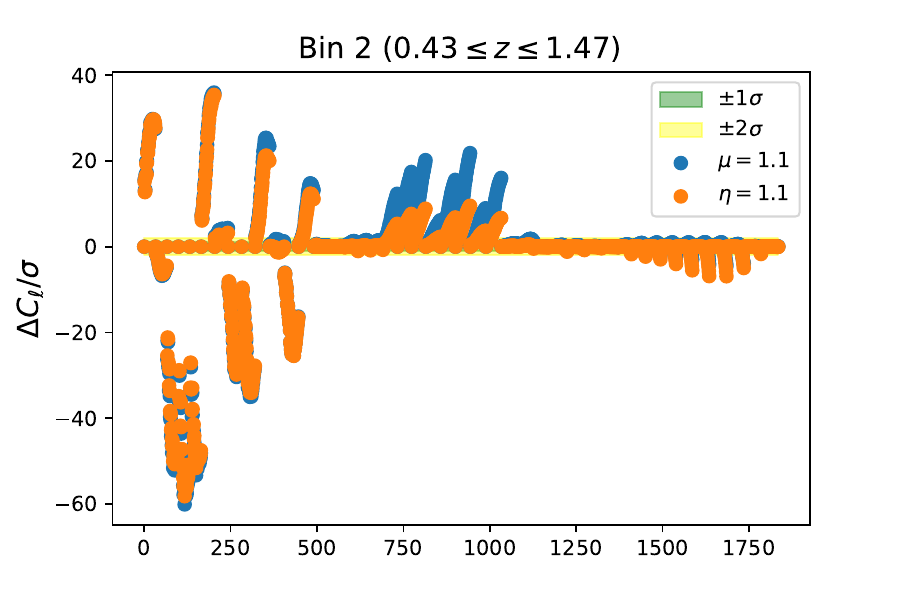}
    \includegraphics[width=0.45\linewidth]{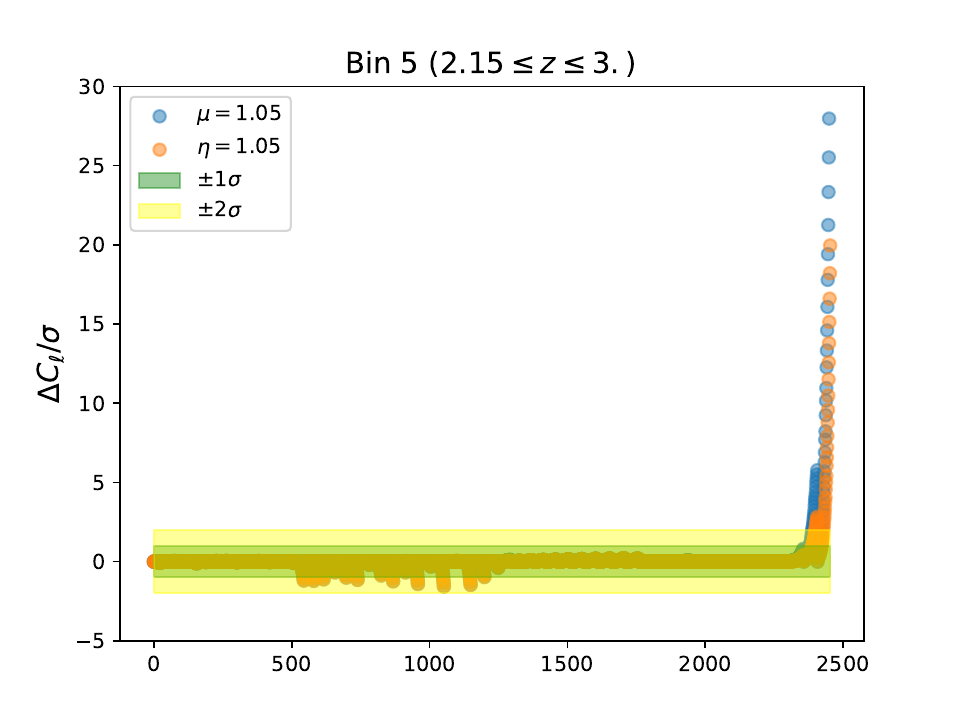}
    \caption{Residuals between the theory prediction from 5\% variation in the MG parameters and the synthetic data vector, shown as $\left(C_\ell^{\rm best}-C_\ell^{\rm data}\right)/\sigma$, where $\sigma$ is the error obtained from the covariance. In the left panel we show the response for the $3\times2$pt data vector when the modification is applied in the lowest redshift bin ($z<0.43$). Both galaxy clustering and cosmic shear exhibit a measurable response, with the signal being more sensitive to variations in $\mu$ than in $\eta$, reflecting the stronger dependence of structure growth on the effective Newtonian coupling. In the right panel we show the response for the $6\times2$pt data vector when the modification is applied in the highest redshift bin ($2.15<z<3$). In this case the $3\times2$pt observables are largely insensitive to the modification and the signal lies within the $1\sigma$ statistical uncertainty, indicating that the large-scale structure probes alone cannot detect such deviations. The dominant response instead arises from the CMB lensing component of the $6\times2$pt data vector, which probes the matter distribution at higher redshifts. As in the low-redshift case, the response is stronger for variations in $\mu$ than in $\eta$.}
    \label{fig:Response}
\end{figure}

In this appendix we perform a set of validation tests of the data vector and modelling pipeline used in the analysis. Since the synthetic data vectors used in this work are generated assuming a fiducial $\Lambda$CDM cosmology, we verify that the pipeline accurately reproduces the input data vector and that the response of the observables to variations in the modified gravity parameters behaves as expected.

We first examine the response of the data vector to small perturbations of the modified gravity parameters $\mu$ and $\eta$. For this test we compute the change in the predicted angular power spectra relative to the fiducial $\Lambda$CDM model and normalize the result by the statistical uncertainty of each data point. The quantity plotted is
\begin{equation}
\frac{\Delta C_\ell}{\sigma_\ell} = \frac{C_\ell^{\rm MG}-C_\ell^{\rm GR}}{\sigma} 
\end{equation}
where $C_\ell^{\rm MG}$ denotes the theoretical prediction in the presence of modified gravity and $C_\ell^{\rm GR}$ corresponds to the fiducial $\Lambda$CDM model used to generate the synthetic data vector and $\sigma$ correpsonds to the 1-sigma error from the covariance.

\begin{figure}
    \centering
    \includegraphics[width=0.7\linewidth]{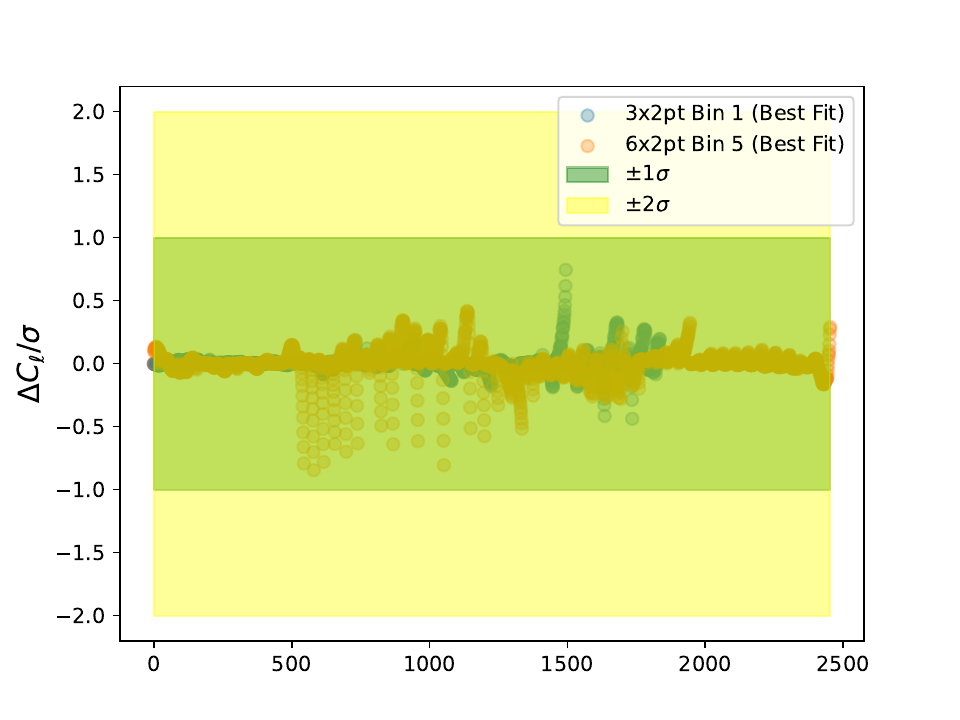}
    \caption{Residuals between the best-fit theory prediction and the synthetic data vector, shown as $\left(C_\ell^{\rm best}-C_\ell^{\rm data}\right)/\sigma$, where $\sigma$ is the error obtained from the covariance. Blue points correspond to the best-fit model from the $3\times2$pt analysis in the lowest redshift bin, while orange points show the best-fit model from the $6\times2$pt analysis in the highest redshift bin. The shaded regions indicate the $\pm1\sigma$ and $\pm2\sigma$ bands expected from the Gaussian covariance. The residuals are consistent with statistical fluctuations and show no systematic trends with multipole, confirming that the modelling pipeline accurately reproduces the synthetic data vector.}
    \label{fig:best_fit}
\end{figure}

Figure~\ref{fig:Response} shows the response of the data vector for two representative cases. In the left panel we consider the $3\times2$pt data vector when the modified gravity parameters are varied in the lowest redshift bin. A 5\% change in either $\mu$ or $\eta$ produces a measurable response in both galaxy clustering and cosmic shear. The response is stronger for variations in $\mu$, reflecting the fact that the growth of structure depends directly on the effective gravitational coupling.

In the right panel we show the corresponding response when the modification is applied in the highest redshift bin. In this case the $3\times2$pt observables are largely insensitive to the modification and the resulting signal lies well within the statistical uncertainty. This reflects the limited sensitivity of large-scale structure probes to modifications of gravity occurring at high redshift. When CMB lensing is included in the $6\times2$pt data vector, however, a detectable response appears due to the sensitivity of CMB lensing to the integrated matter distribution over a broad redshift range extending to $z\sim2$--$5$. As in the low-redshift case, the response is more sensitive to variations in $\mu$ rather than $\eta$.

These results provide a useful physical interpretation of the parameter constraints obtained in the main analysis: low-redshift bins are primarily constrained by galaxy clustering and cosmic shear, while high-redshift bins gain additional sensitivity through the inclusion of CMB lensing.

As a final validation test we examine the residuals between the best-fit model and the synthetic data vector. We compute the residuals between the $\Lambda$CDM data vector and the best-fit prediction from the chains $C_\ell^{\rm best}$.

Figure~\ref{fig:best_fit} shows the residuals for two representative cases: the best-fit model obtained from the $3\times2$pt analysis in the lowest redshift bin and the best-fit model from the $6\times2$pt analysis in the highest redshift bin. In both cases the residuals are consistent with random statistical fluctuations and lie within the expected $\pm1\sigma$ region defined by the Gaussian covariance matrix. No systematic trends with multipole or probe type are observed.

This test confirms that the modelling pipeline accurately reproduces the synthetic data vector and that the likelihood analysis correctly identifies the region of parameter space consistent with the fiducial $\Lambda$CDM cosmology. Together with the response tests discussed above, these results provide a robust validation of the full inference pipeline used in this work.

\section{Sensitivity to Nonlinear Modified Gravity Modelling: Impact of Phenomenological Screening}\label{app:screening}

\begin{figure}
    \centering
    \includegraphics[width=0.45\linewidth]{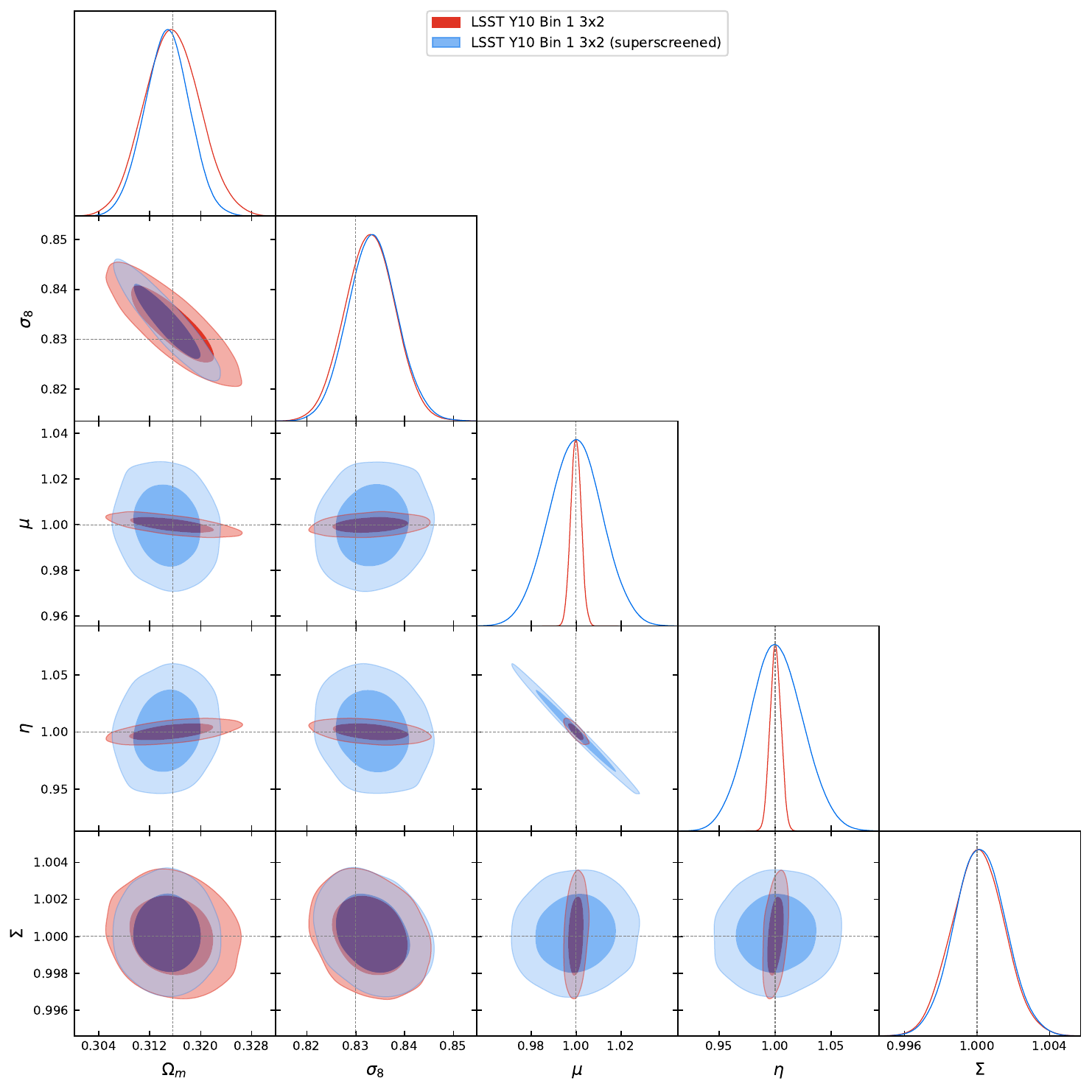}
    \includegraphics[width=0.45\linewidth]{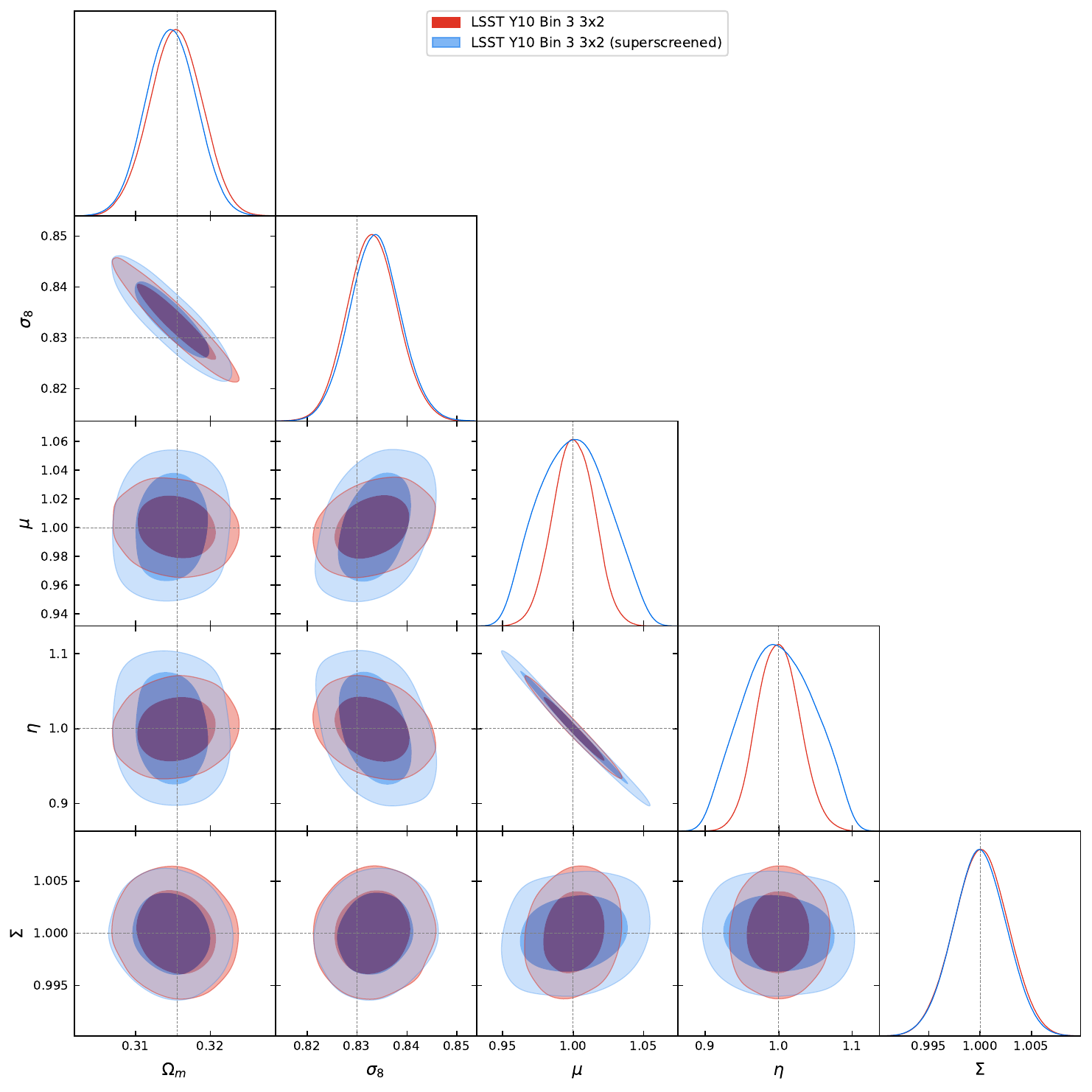}
    \caption{We present the constraints for bin 1 (left panel) and bin 3 (right panel) for the unscreened case (red) and the super-screened case where the non-linear clustering is identical to that in $\Lambda$CDM (blue). We see that non-linear structure growth is instrumental in breaking the $\mu-\eta$ degeneracy, a fact that was noted in previous studies \cite{Srinivasan_2024, Srinivasan2021}.  }
    \label{fig:screening}
\end{figure}

In the main analysis, nonlinear modifications to the matter power spectrum are modelled using the emulator described in Sec.~\ref{sec:emulation}. However, theoretical predictions for nonlinear modified gravity clustering remain uncertain and are susceptible to screening mechanisms that restore general relativity on small scales. To assess the sensitivity of our results to this modelling assumption, we consider a limiting scenario in which nonlinear scales follow the $\Lambda$CDM prediction, while modified gravity affects only the linear regime. This setup is similar to the “super-screened’’ prescription adopted in recent Euclid forecasts of modified gravity constraints \cite{euclid_param_MG_forecast}.

We emphasise that this configuration should not be interpreted as a physically complete implementation of screening. In particular, the prescription modifies the matter power spectrum but does not alter the modification to the lensing potential itself. The purpose of this exercise is therefore not to model screening in a fully consistent way, but rather to provide a robustness test of the impact of nonlinear modified gravity modelling on our inference pipeline.

Figure~\ref{fig:screening} compares parameter constraints obtained using the full nonlinear modified gravity boost with those obtained in this superscreened scenario. We find that the recovered parameters remain consistent within $1\sigma$ across the two cases, indicating that our results are not strongly sensitive to the specific nonlinear prescription adopted. As expected, removing nonlinear modified gravity effects mildly weakens the individual constraints on $\mu$ and $\eta$, since nonlinear clustering contributes to breaking the $\mu$–$\eta$ degeneracy. This behaviour is visible in the right panel, where the $\eta$ constraint approaches the prior boundary in the superscreened case for bin~3.

Importantly, the best-constrained combination $\Sigma=\mu(1+\eta)/2$ remains nearly unchanged between the two setups. This reflects the fact that the constraining power of the $3\times2$pt observables considered here is dominated by quasi-linear scales. We therefore conclude that the main results of this work are robust to reasonable variations in the nonlinear modified gravity modelling, although a fully consistent treatment of screening effects, particularly at the level of ray-traced lensing observables, will ultimately require dedicated simulation studies.

\begin{figure}
    \centering
    \includegraphics[width=0.7\linewidth]{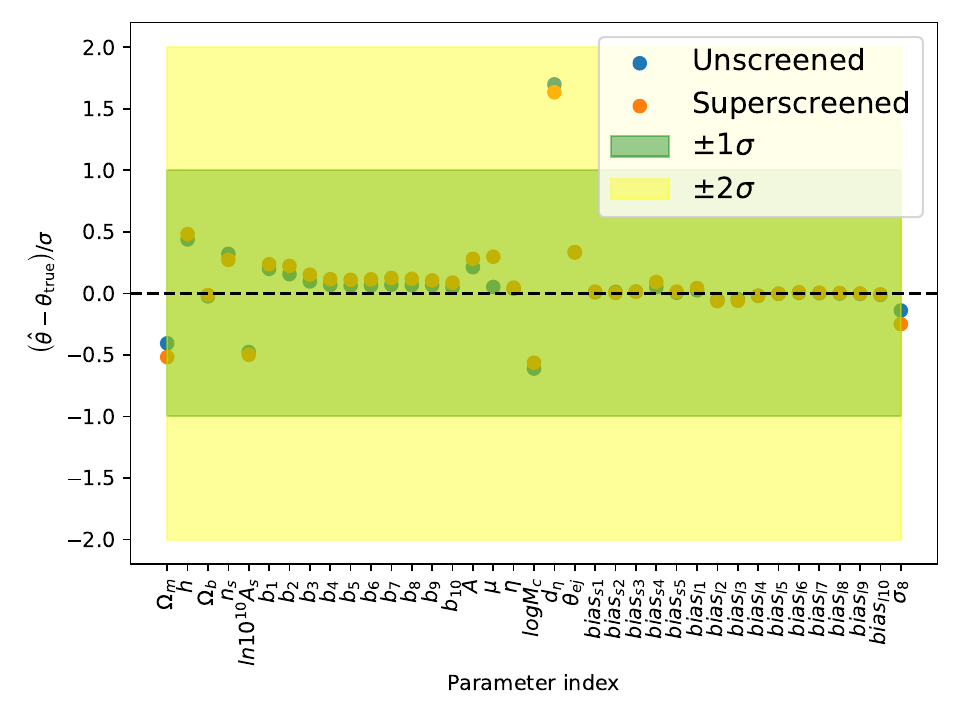}
    \caption{We show the difference between the recovered posterior mean and the injected parameter value, normalized by the posterior standard deviation, for the full set of cosmological, modified-gravity, and nuisance parameters. We show this for the unscreened (blue) and the super-screened (orange) cases. The shaded bands indicate the $\pm1\sigma$ and $\pm2\sigma$ regions. All parameters are recovered within $2\sigma$, with the vast majority lying within $1\sigma$ (except for the baryonic parameter $d_{\rm eta}$ see text for more information), demonstrating that the inference pipeline does not introduce significant biases in the recovered parameters.  }
    \label{fig:screening_recovery}
\end{figure}

As an additional validation of the robustness of the inference pipeline, we perform an injection–recovery test in which mock data generated at the fiducial cosmology are analysed using both the fiducial nonlinear modified-gravity modelling and the superscreened prescription described above. Figure~\ref{fig:screening_recovery} shows the difference between the recovered parameters and their input values for the full set of cosmological, modified-gravity, and nuisance parameters. In all cases the recovered values remain consistent with the input parameters within $1\sigma$, indicating that the choice of nonlinear prescription does not introduce a significant bias in the inferred parameters. The exception to this is the baryonic parameter, $d_{\eta}$. We note that the baryonic feedback parameter $d_{\eta}$ is recovered at the $\sim 1.6\sigma$ level relative to the injected value. This mild shift is due to a volume effect induced by the conservative scale cut in this analysis. We check to see that this shift vanishes for a $k_{\rm cut} = 1\,h\,{\rm Mpc}^{-1}$. As a result, we remark that this actually indicates the power of the $k_{\rm cut}$ method in that it ensures that the important cosmological parameters are not biased due to baryonic effects. This further supports the interpretation of the superscreened setup as a robustness test of the modelling assumptions rather than as a physically complete description of screening effects.

\end{document}